\documentclass[useAMS,usenatbib]{mn2e}

\usepackage{amsmath}
\usepackage{graphicx}
\usepackage{txfonts}
\usepackage{natbib}
\pdfoutput=1

\newcommand{\teff}{T_{\rm{eff}}}
\newcommand{\logg}{\log g}
\newcommand{\feh}{\rm{[Fe/H]}}

\newcommand{\kepler}{{\it Kepler~}}
\newcommand{\stromgren}{Str\"omgren~}
\def\dnu{\Delta\nu}
\def\num{\nu_\mathrm{max}}
\def\msun{\rm M_{\odot}}

\title[Age stratigraphy of the Milky Way disc]{Measuring the vertical age
  structure of the Galactic disc using asteroseismology and SAGA\thanks{http://www.mso.anu.edu.au/saga}}

\author[Casagrande et al.]{\parbox{18cm}{
  L. Casagrande$^{1,2}$\thanks{Stromlo Fellow}\thanks{luca.casagrande@anu.edu.au},
  V. Silva Aguirre$^{3,2}$,
  K.J. Schlesinger$^1$,
  D. Stello$^{4,3,2}$,
  D. Huber$^{4,5,3,2}$,
  A.M. Serenelli$^{6,2}$,
  R. Sch\"onrich$^7$,
  S. Cassisi$^8$,
  A. Pietrinferni$^8$,
  S. Hodgkin$^9$,
  A.P. Milone$^1$,
  S. Feltzing$^{10}$,
  M. Asplund$^1$}\vspace{0.3cm}\\
$^1$ Research School of Astronomy \& Astrophysics, Mount Stromlo 
   Observatory, The Australian National University, ACT 2611, 
   Australia\\
$^2$ Kavli Institute for Theoretical Physics, University of California,
   Santa Barbara, CA 93106, USA\\   
$^3$ Stellar Astrophysics Centre, Department of Physics and 
   Astronomy, Aarhus University, Ny Munkegade 120, DK-8000 
   Aarhus C, Denmark\\   
$^4$ Sydney Institute for Astronomy (SIfA), School of Physics, 
   University of Sydney, NSW 2006, Australia\\   
$^5$ SETI Institute, 189 Bernardo Avenue, Mountain View, CA 94043,
   USA\\
$^6$ Instituto de Ciencias del Espacio (ICE-CSIC/IEEC) Campus UAB,
   Carrer de Can Magrans, s/n 08193 Cerdanyola del Vallés\\
$^7$ Rudolf-Peierls Centre for Theoretical Physics, University of 
    Oxford, 1 Keble Road, OX1 3NP, Oxford, United Kingdom\\
$^8$ INAF-Osservatorio Astronomico di Collurania, via Maggini, 
    64100 Teramo, Italy\\
$^9$ Institute of Astronomy, Madingley Road, Cambridge CB3 0HA, UK\\
$^{10}$ Lund Observatory, Department of Astronomy \& Theoretical 
                  Physics, Box 43, SE-22100, Lund, Sweden}

\begin{document}

\date{Received; accepted}

\maketitle

\begin{abstract}

  The existence of a vertical age gradient in the Milky Way disc has been
  indirectly known for long. Here, we measure it directly for the first
  time with seismic ages, using red giants observed by {\it Kepler}. We use
  Str\"omgren
  photometry to gauge the selection function of asteroseismic targets,
  and derive colour and magnitude limits where giants with measured oscillations
  are representative of the underlying population in the field. Limits
  in the 2MASS system are also derived. We lay out a method to assess and
  correct for target selection effects independent of Galaxy models. We find
  that low mass, i.e.~old red giants dominate at increasing Galactic heights,
  whereas closer to the Galactic plane they exhibit a wide range of ages and
  metallicities.
  Parametrizing this as a vertical gradient returns approximately
  $4$\,Gyr kpc$^{-1}$ for the disc we probe, although with a large
  dispersion of ages at all heights. The ages of stars show a
  smooth distribution over the last $\simeq10$\,Gyr, consistent with a mostly
  quiescent evolution for the Milky Way disc since a redshift of about 2. We
  also find a flat age-metallicity relation for disc stars. Finally, we
  show
  how to use secondary clump stars to estimate the present-day intrinsic
  metallicity spread, and suggest using their number count as a new proxy for
  tracing the aging of the disc. This work highlights the power of
  asteroseismology for Galactic studies; however, we also emphasize the need
  for better constraints on stellar mass-loss, which is a major source of
  systematic age uncertainties in red giant stars.

\end{abstract}


\section{Introduction}\label{sec:intro}

A substantial fraction of the baryonic matter of the Milky Way is contained in 
its disc, where much of the evolutionary activity takes place. Thus, knowledge  
of disc properties is crucial for understanding how galaxies form and evolve. 
Late-type Milky Way-like galaxies are common in the local
universe. However, we can at best measure integrated properties for
external galaxies, while the Milky Way offers us the unique
opportunity to study its individual baryonic components.

Star counts have revealed that the disc of the Milky Way is best described 
by two populations, one with shorter and one with longer scale-heights, dubbed 
the ``thin'' and the ``thick'' disc \citep[e.g.,][]{gr83,juric08}. This double 
disc behaviour is also inferred from observations of edge-on galaxies, where 
the thick disc appears as a puffed up component extending to a larger height 
above a sharper thin disc \citep[e.g.,][]{burstein79,vdKS81_891,yd06}.
Although it is usually possible to fit the vertical density/luminosity profile
of late-type galaxies as a double-exponential profile, its interpretation is
still a matter of debate.
In particular, it is unclear if the thin and thick disc in the Milky Way 
are real, separated structural entities, or not \citep[e.g.,][]{norris87,nn91,
  sb09b,brh12}.
These different interpretations on disc's decomposition underpin much of the
theoretical framework for understanding its origin and evolution. Models in
which the thick disc is formed at some point during the history of the Galaxy
via an external mechanism (in particular 
accretion and/or mergers) best fit the picture in which the thin and the thick 
disc are real separated entities \citep[e.g.,][]{cmg97,abadi03,bkg04,villa08,
kbz08,scanna09}. This scenario acquired momentum in the framework of cold dark 
matter models, where structures (and galaxies) in the universe form 
hierarchically \citep[e.g.,][]{wr78,sz78}. Thus, the growth of a spiral galaxy
over cosmic time would be responsible for puffing up the disc, also ``heating''
the kinematics of its stars.
In contrast, internal dynamical evolution \citep[primarily in the form of 
radial mixing e.g.,][]{sb09a,sb09b,lrd11} favours the scenario in which the 
thick disc is the evolutionary end point of an initially pure thin disc, 
without requiring a heating mechanism. Internal sources of dynamical 
disc-heating, e.g. from giant molecular clouds or clump-induced stellar
scattering, may also contribute to thick disc formation \citep[e.g.,][]
{hf02,bem09}.
Although merger events can happen at early times, in this picture the 
formation of the Galaxy is mostly quiescent.

As is often the case, the real --yet unsolved-- picture of galaxy formation 
is more complicated than the simplistic sketches drawn above. The latest
numerical simulations indicate that hierarchical accretion occurring at
early times can imprint a signature of hot kinematic and roundish structures.
Other processes then factor into the following more quiescent phases, 
characterized by the formation of younger disc stars in a more flattened,
rotationally supported configuration \citep[e.g.,][]{genzel06,bem09,aumer10,
  house11,forbes12,stinson13,bird13}. Importantly, 
high-redshift observations suggest that, for galaxies in the Milky Way mass 
range, this might not happen in an inside-out fashion \citep{vD13}.

Historically, the study of chemical and kinematic properties of stars in the
(rather local) disc, has been used to shed light on these different formation
scenarios. 
Thin disc stars are observed to be on average more metal-rich and less
alpha-enhanced than thick disc stars \citep[e.g.,][]
{edvardsson93,chen00,reddy03,fuhrmann08,bfo14}. Due to stellar 
evolutionary timescales the enrichment in alpha elements happens relatively
quickly \citep[e.g.][]{tinsley79,mg86}. Thus, for the bulk of local disc
stars it is customary to interpret this chemical distinction into an
age difference (but see \citealt{chi15} and \citealt{martig15} for the
possible existence of local outliers). In this picture the thick disc would
be the result of some event in the history of the early Galaxy and thus
metal-poor and alpha-enhanced. This interpretation however has been recently
challenged by the observational evidence that alpha-enhanced thick disc stars
may also extend to super-solar metallicities \citep[e.g.,][]
{fb08,c11,tbe11,bfo14}. This can be explained if --at least some of-- the
thick disc is composed of stars originating from the inner Galaxy, where
the chemical enrichment happened faster.  
In terms of kinematics, thin disc stars are cooler (i.e.~with smaller vertical
velocity with respect to the Galactic plane) and have higher Galactic
rotational velocity compared to thick disc stars then referred to as
kinematically hot. Low rotational velocities (due to larger asymmetric drift)
imply higher velocity dispersion for thick disc stars, which then point to
older ages, either born hot or heated up. In fact, the age-velocity
dispersion relation has long been known to indicate the existence of a
vertical age gradient \citep[e.g.][]{vH60,mayor74,w77,holmberg07}: its direct
measurement is the subject of the present study.

The dissection of disc components based only on chemistry and kinematic is far 
from trivial \citep[e.g.,][]{sb09b}. In this context, stellar ages are 
expected to provide an additional powerful criterion. Also, age cohorts are 
easier to compare with numerical simulations than chemistry based 
investigations, bypassing uncertainties related to the implementation of the 
chemistry in the models. From the observational point of view however, even 
when accurate astrometric distances are available to allow comparison of stars 
with isochrones, the derived ages are still highly uncertain, and statistical 
techniques are required to avoid biases \citep[e.g.,][]{pe04,jl05,s13}. 
Furthermore, isochrone dating is meaningful only 
for stars in the turnoff and subgiant phase, where stars of different ages 
are clearly separated on the H-R diagram. This is in contrast, for example, to 
stars on the red giant branch, where isochrones with vastly different 
ages can fit observational data such as effective temperatures, metallicities,
and surface gravities equally well within their errors. As a 
result, so far the derivation of stellar ages has been essentially limited to
main-sequence F and G type stars with {\it Hipparcos} parallaxes, i.e. around
$\sim 100$~pc from the Sun \citep[e.g.,][]{fhh01,bensby03,nordstrom04,haywood08,c11}.
All these studies agree on the fact that the thick disc is older than the thin 
disc. Yet, only a minor fraction of stars in the solar neighbourhood belong 
to the thick disc. 

It is now possible to break this impasse thanks to asteroseismology. In 
particular, the latest spaceborne missions such as {\it CoRoT} \citep{corot06} 
and {\it Kepler}/K2 \citep{gilli10,k2} allow us to robustly measure global 
oscillation frequencies in several thousands of stars, in particular red 
giants, which in turn make it possible to determine fundamental physical 
quantities, including radii, distances and masses. Most importantly, once a 
star has evolved to the red giant phase, its age is determined to good 
approximation by the time spent in the core-hydrogen burning phase, and this is 
predominantly a function of the stellar mass. Although mass-loss can partly 
clutter this relationship, as we will discuss later in the paper, to 
first approximation the mass of a red giant is a proxy for its 
age. In addition, because of the intrinsic luminosity of red giants, they can 
easily be used to probe distances up to a few kpc \citep[e.g.][]{miglio13a}. 

This has profound impact for Milky Way studies, and in fact synergy with 
asteroseismology is now sought by all major surveys in stellar and Galactic 
archaeology \citep[e.g.][]{pin14,galah}. With similar motivation, we have 
started the \stromgren survey 
for Asteroseismology and Galactic Archaeology \citep[SAGA,][hereafter Paper 
I]{c14a} which so far has derived classic and asteroseismic stellar 
parameters for nearly 1000 red giants with measured seismic oscillations in the 
\kepler field. In this paper we derive stellar ages for the entire SAGA
asteroseismic sample, and use them to study the vertical age 
structure of the Milky Way disc. Our novel approach uses the power of
seismology to address thorny issues in Galactic evolution, such as the
age-metallicity relation, and to provide in situ measurements of stellar ages
at different heights above the Galactic plane, at the transition between the
thin and the thick disc. The study of the vertical metallicity structure of
the disc with SAGA will be presented in a companion paper by \cite{sch14}.

The paper is organized as follows: in Section \ref{sec:SAGA} we review
the SAGA survey, and present the derivation of stellar ages for the seismic
sample. In Section \ref{sec:stat} we investigate the selection function of
the {\it Kepler} satellite, and identify the colour and magnitude intervals
within which the asteroseismic sample is representative of the underlying
stellar population in the field. This allows us to define clear selection
criteria, which are then used in Section \ref{sec:grad} to derive vertical
mass and age gradients. We provide raw gradients, i.e.~obtained by
simply fitting all stars that pass the selection criteria in Section
\ref{sec:grad_method}. The biases introduced by our target selection criteria
are also assessed, and gradients corrected for these effects are presented
in Section \ref{sec:grad_bias}. The implications of the age-metallicity
relation, and of the age distribution of red giants to constrain the 
evolution of the Galactic disc are discussed in Section \ref{sec:agemet}.
In Section \ref{sec:rc2} we suggest using secondary 
clump stars as age candles for Galactic Archaeology. Finally, we
conclude in Section \ref{sec:end}.

\section{The SAGA}\label{sec:SAGA}

The purpose of the SAGA is to uniformly and homogeneously observe stars in the 
\stromgren $uvby$ system across the \kepler field, in order to transform it 
into a new benchmark for Galactic studies, similar to the solar neighbourhood. 
Details on survey rationale, strategy, observations and data reduction are 
provided in Paper I, and here we briefly summarize the information 
relevant for the present work. So far, observations of a stripe centred at 
Galactic longitude $l \simeq 74^{\circ}$ and covering latitude 
$8^{\circ} \lesssim b \lesssim 20^{\circ}$ have been reduced and analyzed. This 
geometry is particularly well suited to study vertical gradients in the 
Galactic disc. 

The \stromgren $uvby$ system \citep{stromgren63} was designed for the 
determination of basic stellar parameters \citep[see e.g.,][and references 
therein]{arnadottir10}. Its $y$ magnitudes are defined to be essentially the 
same as the Johnson $V$ \citep[e.g.,][]{bessell05}, and in this work we will
refer to the two interchangeably.

SAGA observations are conducted with the Wide Field Camera on the $2.5$-m 
Isaac Newton Telescope (INT), which in virtue of its large field of view and 
pixel size is 
ideal for wide field optical imaging surveys. The purpose of our survey is to 
obtain good photometry for {\it all} stars in the magnitude range
$10 \lesssim y \lesssim 14$, where most targets were selected to measure
oscillations with {\it Kepler}. This requirement can be easily achieved with
short exposures on a $2.5$-m 
telescope; indeed, all stars for which \kepler measured oscillations are 
essentially detected in our survey (with a completeness $ \gtrsim 95$ per cent, 
see Paper I). SAGA is magnitude complete to about $y \simeq 16$~mag, thus
providing 
an unbiased, magnitude-limited census of stars in the Galactic stripe observed. 
Stars are still detected at fainter magnitudes ($y \lesssim 18$), although 
with increasingly larger photometric errors and incompleteness, totalling some 
$29,000$ objects in the stripe observed so far. Thus, we can build two samples 
from our observations. First, a magnitude complete and unbiased photometric 
sample down to $y \simeq 16$~mag, which we refer to as the full photometric 
catalog. Second, we extract a subset of 989 stars which have oscillations 
measured by {\it Kepler}, dubbed the asteroseismic catalog. Note that the
stars for which \kepler measured oscillations were selected in a non trivial
way. By comparing the properties of the asteroseismic and full photometric
catalogs we can assess the \kepler selection function (Section \ref{sec:stat}).

Before addressing the \kepler selection function, we briefly recall the 
salient features of the full photometric and asteroseismic catalog. For 
the asteroseismic catalog we also derive stellar ages, and discuss their 
precision. 

\subsection{Full photometric sample}

The full photometric catalog provides $uvby$ photometry for several thousands 
of stellar sources down to a magnitude completeness limit of $y \simeq 16$. 
The asteroseismic catalog is a subset of the 
targets in the full photometric catalog, and thus the data reduction and 
analysis is identical for all stars in SAGA. In this work, for both the 
asteroseismic and the full photometric sample (or any sub-sample extracted 
from them) we use only stars with reliable photometry in all $uvby$ bands 
(Pflg={\tt 0} flag, see also figure  4 in Paper I). This requirement excludes 
stars whose errors are larger than the ridge-lines defined by the bulk of 
photometric errors, as customarily done in photometric analysis, and does not 
introduce any bias for our purposes. 
Furthermore, we have verified that the fraction of stars excluded as a function 
of increasing $y$~mag is nearly identical for both the asteroseismic and the 
full photometric sample. This is true for the two samples as a whole, or when 
restricting them only to giants.
A flag also identifies stars with reliable $\feh$, 
i.e.~those objects for which the \stromgren metallicity calibration is used 
within its range of applicability (Mflg={\tt 0}). This flag automatically 
excludes stars with $\feh \geqslant 0.5$, where such high values could be the 
result of extrapolations in the metallicity calibration and/or stem from 
photometric and reddening errors. Again, this limit is not expected to 
introduce any bias given the paucity (or even the non-existence, see e.g.,
\citealt{taylor06}) of star more metal-rich than $0.5$~dex in nature
(i.e.~those present in our catalog are likely flawed).
Later in the paper, we will use only stars with both Pflg and Mflg equal to 
zero to constrain the \kepler selection function and to study the vertical 
stellar mass and age struture in the Galactic disc. 
\begin{figure*}
\begin{center}
\includegraphics[scale=0.14]{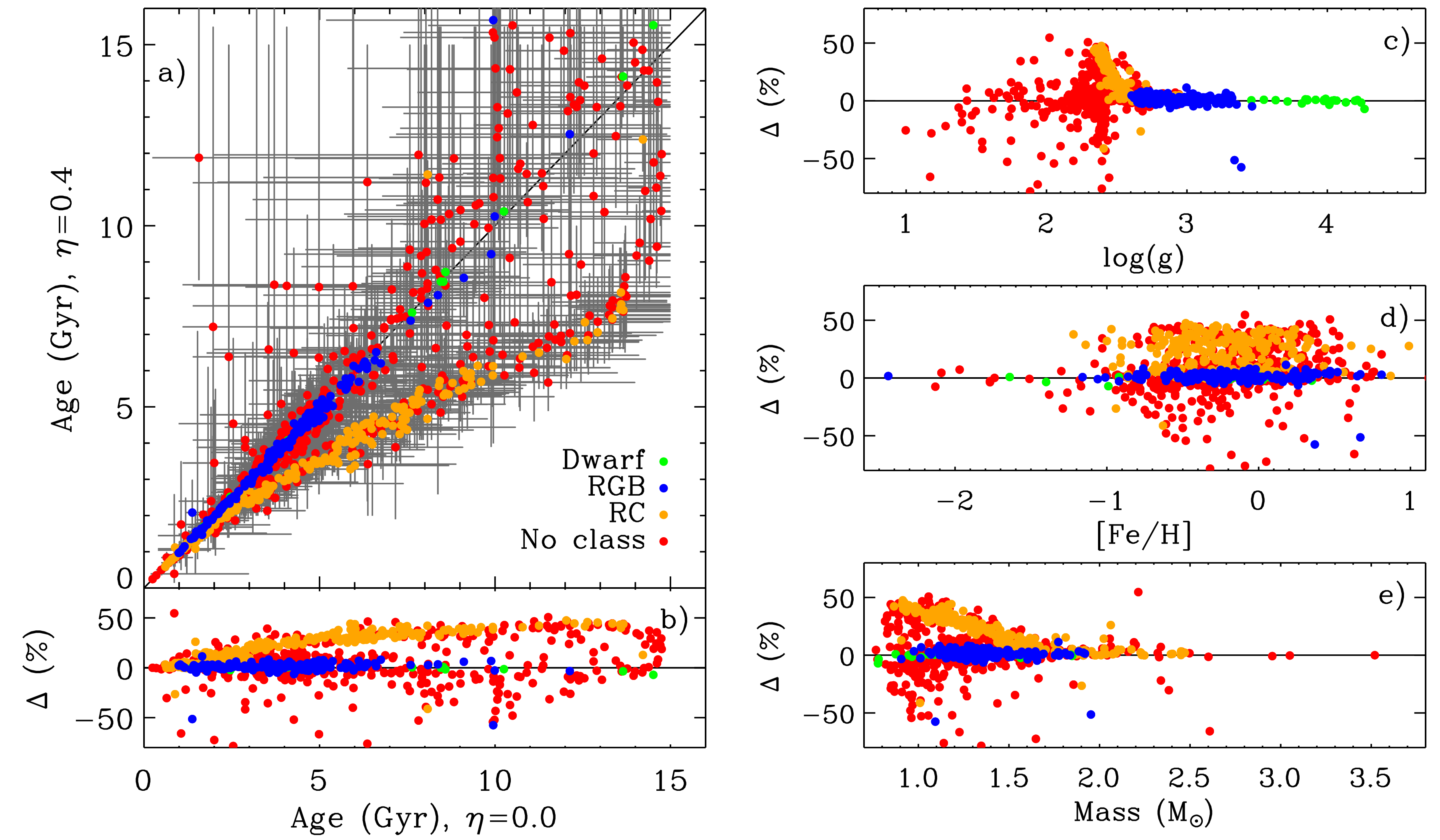}
\caption{{\it Panel} a): comparison between ages obtained with mass-loss 
(vertical axis, $\eta=0.4$), and without mass-loss (horizontal axis). Colours 
identify stars with different seismic classification, as labelled (see also 
Paper I). Error bars are formal uncertainties, only. {\it Panel} b) to e):
same as above, but showing the fractional age difference as function of
different parameters.}
\label{f:ages}
\end{center}
\end{figure*}

\subsection{Asteroseismic sample}\label{subsec:as}

The SAGA asteroseismic catalog consists of 989 stars identified by
cross-matching our \stromgren observations with the dwarf sample of
\cite{chap14} and the $\simeq 15,000$ giants from the \kepler Asteroseismic
Science Consortium (KASC, \citealt{stello13,h14}). Within SAGA, a novel
approach is developed to couple classic and asteroseismic stellar parameters:
for each target, the photometric effective temperature and metallicity,
together with the asteroseismic mass, radius, surface gravity, mean density and
distance are computed \citep{c10,c14a,vsa11,vsa12}. A detailed assessment of
the uncertainties in these parameters is given in Paper I. For a large
fraction of objects, evolutionary phase classification identifies whether a
star is a dwarf (labelled as ``Dwarf'', 23 such stars in our sample), is
evolving along the
red giant branch (``RGB'') or is already in the clump phase (``RC''). It was
possible to robustly distinguish between the last two evolutionary phases for
427 stars, whereas for the remaining giants no classification is available
(``NO''). In this paper, we refer to all stars classified as ``RGB'', ``RC'',
or ``NO'' as red giants.

\subsection{Asteroseismic ages}\label{sec:ages}

With the information available for each asteroseismic target it is rather 
straightforward to compute stellar properties. As described in Paper~I, we 
apply a Bayesian scheme to sets of BaSTI isochrones\footnote{http://www.oa-teramo.inaf.it/BASTI} \citep{pie04,pie06}. Flat priors are assumed for ages
and metallicities over the entire grid of BaSTI models, meaning that at all
ages, all metallicities are equally possible. A \citet[][]{salpeter55} 
Initial Mass Function (IMF, $\alpha=-2.35$) is also used \citep[see details
in][]{vsa15}.
The adopted asteroseismic stellar parameters are derived using non-canonical
BaSTI models with no mass-loss, but we explore the effect of varying some of
the BaSTI prescriptions as described further below. As input parameters we
consider the two global asteroseismic parameters $\dnu$ and $\num$ and the
atmospheric observables $\teff$ and $\feh$. The information on the
evolutionary phase (``RGB'', ``RC'') is included as a prior when available,
otherwise the probability that a star belongs to a given evolutionary
status is determined by the input observables, the adopted IMF and the
evolutionary timescales. The median and 68 per cent confidence levels of the
probability distribution function determine the central value and (asymmetric)
uncertainty of ages, which following the terminology of Paper I we refer to as
formal uncertainties. 
\begin{figure*}
\begin{center}
\includegraphics[scale=0.14]{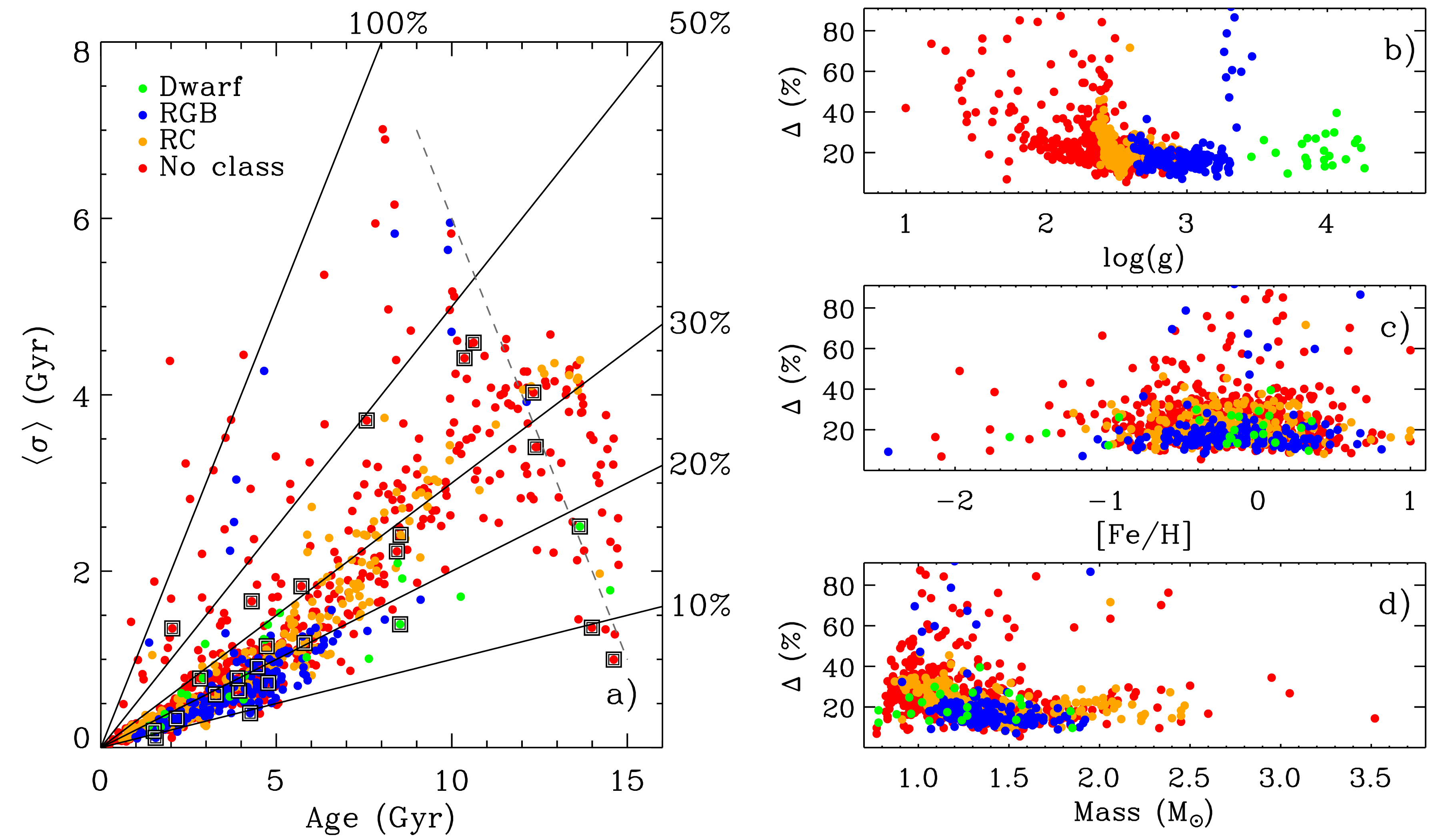}
\caption{{\it Panel} a): final global age uncertainties 
$\langle \sigma \rangle$ (as defined in the text) for stars with different 
seismic classification plotted as function of their ages. Squares highlight 
stars with $\feh < -1$, whereas continuous ridge-lines mark uncertainties
between 10 and 100 per cent. Dashed line is the maximum uncertainty 
formally possible at old ages because of the sharp cut imposed at 15~Gyr.
{\it Panel} b) to d): fractional age uncertainties as function of 
different parameters.}
\label{f:eage}
\end{center}
\end{figure*}

Overshooting in the main-sequence phase can significantly change the turn-off
age of a star and is therefore important for our purposes. In order to assess
its impact, in the present analysis we explore the effect of using BaSTI
isochrones
computed from stellar models not accounting for core convective overshoot
during the central H-burning stage (dubbed canonical) as well as isochrones
based on models including this effect (dubbed non-canonical, and adopted as
reference). We note
that all sets of BaSTI isochrones take into account semiconvection during the
core helium-burning phase. The BaSTI theoretical framework for mass-loss
implements the recipe of \citet{rei75}:
\begin{equation}\label{eq:reim}
\frac{dM}{dt}=\eta \, 4\times10^{-13} \frac{L}{gR} \, \left[\frac{\msun}{{\rm yr}}\right]\,, 
\end{equation}
where $\eta$ is a (free) efficiency parameter that needs to be constrained by 
observations \citep[see e.g. the recent analyses by][]{McDZ15,heyl15}. As we
discuss
further below, mass-loss efficiency can have a considerable impact on the
parameters relevant for the present analysis. Sets of BaSTI stellar models
have been computed for different values of $\eta$; we explore its effects by
using the no mass-loss isochrones ($\eta=0$) as our reference set and compare
to the stellar properties derived with the $\eta=0.4$ set.

As we did for the other seismic parameters, we pay particular attention to 
derive realistic uncertainties for our age estimates. The computation of ages 
for red giant stars heavily rely on the knowledge of their masses. Here, the 
mass of a star is essentially fit (with a grid-based Bayesian scheme) using 
scaling relations based on the observed $\Delta\nu$ and $\nu_{\rm{max}}$, and 
thus the mass is rather independent of the models adopted. Paper I
demonstrated that for most stellar parameters, assuming a very efficient
mass-loss 
($\eta=0.4$) or neglecting core overshooting during the H-core burning 
phase affects the results significantly less than the formal uncertainty of 
the property under consideration. The same, however, does not hold for ages. 
Although the mass of a red giant is a useful proxy for age, it is important to
distinguish between the initial mass which sets the evolutionary lifetime of
a star, and the present-day (i.e.~actual) stellar mass, which is derived from
seismology. Thus, when inferring an age using the actual stellar mass, the
value derived will depend on the past history of the star, whether or not
significant mass-loss has occurred during its evolution.

There are very few observational constraints on mass-loss. Open clusters in 
the \kepler field suggest a value of $\eta$ between $0.1$ and $0.3$ 
\citep{mbs12} for solar metallicity stars of masses $\sim$1.2--1.5~$\msun$;
observations of globular clusters reveal that mass-loss seems to be episodic
and increasingly important when ascending the red giant branch
\citep[see e.g.,][]{origlia14}, with recent studies suggesting a very
inefficient mass-loss during this phase \citep{heyl15}. 
For the asteroseismic sample we also derive ages using BaSTI isochrones 
with $\eta =0.4$. This value corresponds to an efficient mass-loss process, and
it is often used e.g., to reproduce the mean colours of horizontal branch stars
in Galactic globular clusters, although this morphological feature is affected
by other, also poorly constrained, parameters \citep[e.g.,][]{catelan09,gcb10,mmd14}. By deriving ages with both $\eta=0$ and $0.4$, we can compare two
extreme cases and derive a conservative estimate of the age uncertainty
introduced by mass-loss. 

The comparison of ages derived with and without mass-loss is done in Figure 
\ref{f:ages}. Ages of dwarf stars are obviously unaffected by mass-loss, and 
the same conclusion holds for stars with ``RGB'' classification. 
It must be noticed that the distinction between ``RGB'' and ``RC'' is 
based on the average spacing between mixed dipole modes, and this measurement 
largely depends on the frequency resolution which smoothes over the spacing 
\cite[e.g.,][]{Bedding:2011il}.
A clear identification of ``RGB'' stars is thus possible for 
$\logg \gtrsim 2.6$ i.e.~on the lower part of the giant branch, where 
mass-loss turns out to be of little or no importance in the Reimers' 
formulation. This explains the weak dependence of ``RGB'' ages on mass-loss. 
The effect of mass-loss increases when moving to lower gravities, and it is 
most dramatic for stars in the clump phase.

Isochrones including mass-loss return younger ages than those without 
mass-loss; this can be easily understood since a given mass --seismically 
inferred-- will correspond to a higher initial mass in case of mass-loss, and
thus evolve faster to its presently observed value. From
Equation~\ref{eq:reim} it can be seen that the rate of mass-loss has an
inverse dependence on mass. This implies a decreasing 
importance of mass-loss for increasing stellar mass. This is evident in Figure 
\ref{f:ages}e, where only masses below about $1.7 M_{\odot}$ are significantly 
affected by mass-loss.
The fractional differences shown in Figure \ref{f:ages} deserve an obvious 
-yet important- word of caution. We define the reference ages as those 
without mass-loss. Within this context, a fractional difference of e.g., 50
per cent 
means that age estimates decrease by half when we factor in mass-loss. 
Should the same difference be computed using $\eta=0.4$ as reference, ages
from mass-loss should then be increased by twice, implying a 100 per cent
change in this example. 

In addition to mass-loss, we have also tested the effect of canonical and 
non-canonical models (for a given $\eta$). The difference is negligible 
above 3\,Gyr, with differences of a few per cent or less, while at younger ages 
(i.e.~for masses above $\simeq 1.4 M_{\odot}$) the effect can amount to a few 
hundreds Myr, thus translating in age differences of few tens of per cent for 
the youngest stars. The reason for this is that in this mass range, the 
inclusion of overshooting in the main-sequence phase plays a significant role 
in the turn-off age. Although part of this difference is compensated by a 
quicker evolution in the subgiant phase for stars with smaller helium cores 
\citep[i.e. with no overshooting, see][]{maeder74}, the effect remains in more 
advanced stages.

To determine our final and global uncertainties on ages we adopt the same 
procedure used for other seismic parameters, but also account for the 
uncertainties related to mass-loss and the use of (non)-canonical models (see 
Paper I for details on the GARSTEC grid and Monte-Carlo approach discussed 
below). Briefly, we add quadratically to the formal asymmetric 
uncertainties obtained from our $\eta=0$ non-canonical BaSTI reference models 
half the difference between these results and the ones obtained with 
{\it i)} the GARSTEC grid, {\it ii)} the Monte-Carlo approach, {\it iii)} 
implementing mass-loss with $\eta=0.4$ and {\it iv)} using BaSTI canonical 
models. In most cases the uncertainties listed in {\it i)} to {\it iv)}
dominate over the asymmetric formal uncertainty. For plotting purposes we use 
$\langle \sigma \rangle$ defined as the average of the (absolute) value of the 
upper and lower age uncertainty. Figure \ref{f:eage} shows both the absolute 
and relative age uncertainty of each star in our sample, along with their 
dependence on $\logg, \feh$ and mass.

For most of the stars, the age uncertainty is between 10 and 30 per cent. When 
restricting to gravities higher than $\logg \simeq 2.6$, uncertainties of
order 20 per cent are common. The lower part of the red giant branch is where
the effect of mass-loss is weak for stars ascending it, and where seismic 
classification is able to separate ``RGB'' from ``RC'' stars. There is only a 
handful of ``RGB'' stars with uncertainties larger than about 30 per cent:
those are located at the base of the red giant branch and have $\Delta\nu$,
but not 
$\nu_{\rm{max}}$ measurements, explaining their larger errors. For dwarfs, our
age uncertainties are also consistent with the results of \cite{chap14}, who
found a median age uncertainty of 25 per cent  when having good 
constraints on $\teff$ and $\feh$. For the 20 dwarfs we have in common with 
that work, which span an interval of about 10\,Gyr, the mean age difference 
is 1\,Gyr with a scatter of 3\,Gyr. The largest differences occour for the most 
metal-poor stars, and the stars having Pflg and Mflg different from zero. These 
discrepancies likely arise from \cite{chap14} assuming a constant 
$\feh=-0.2$ for all targets, but also our flagged stars might have less 
reliable metallicities.

At the oldest ages, formal uncertainties decrease because of the cut imposed 
at 15\,Gyr (this is true for the upper uncertainty on ages, but obviously also 
the average $\langle \sigma \rangle$ is affected, see Figure \ref{f:eage}). 
Notice that our global uncertainties (which include the effect of different 
models and mass-loss assumptions) partly blur this limit. We also remark that 
the accuracy of asteroseismic masses (and thus ages) obtained from scaling 
relations is still largely unexplored, especially in giants \cite[see e.g.,][]
{miglio13b}. There are also indications that in the metal-poor regime
($\feh \lesssim -1$) scaling relations might overestimate stellar masses by
$15-20$ per cent \citep{eej14}, thus returning ages systematically younger by
more than 60 per cent \citep[see e.g.,][]{andressa12}. In absence of a more
definitive assessments on the limits of the scaling relations, 
this source of uncertainty has not been included in our error budget. Our 
metal-poor stars are highlighted in Figure \ref{f:eage}, and they cover the 
entire age range (i.e.~we do also have metal-poor old stars) with formal 
uncertainties between 20 and 30 per cent (should scaling relations for
metal-poor stars be trusted).

\subsubsection{Asteroseismic ages: reality check}

\begin{figure}
\begin{center}
\includegraphics[scale=0.5]{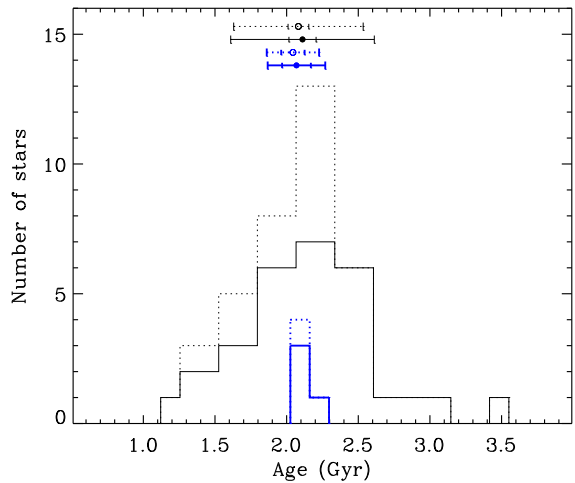}
\caption{Age histogram for radial-velocity single members of the open cluster
  NGC\,6819, selected according to the seismic membership of \citealt{stello11}.
  Thin black dotted line indicates all stars belonging to the cluster, without
  any further pruning. Thin continuous black line is when restricting the
  sample to stars with good Mflg and Pflg. Thick blue lines (dotted and
  continuous) are when further restricting to ``RGB'' stars. Circles in
  the upper part of the plot identify the mean value of each histogram,
  together with its standard deviation (outer bar) and the standard deviation
  of the mean (inner bar).}
\label{f:6819}
\end{center}
\end{figure}

As for the other seismic parameters in Paper I, the solar-metallicity open
cluster NGC\,6819 
offers an important benchmark to check our results. In Figure \ref{f:6819} we 
show the age distribution of its cluster members, from using all seismic
members \citep{stello11} to only a subset of them with the best \stromgren
photometry and seismic evolutionary phase classification. 
We recall that for each star belonging to the cluster, we use its own 
metallicity rather than imposing the mean cluster $\feh$ for all its members.
Requiring good Mflg and Pflg does not seem to reduce the scatter, and thus 
improve the quality of the ages. This is partly expected: although our 
Bayesian scheme fits a number of observables, the main factor in determining
ages is the stellar mass, which is mostly constrained by the asteroseismic
observables. 
More crucial in improving the age precision is to select ``RGB'' stars only, 
from which we derive a mean (and median) age of $2.0 \pm 0.2$\,Gyr for this 
cluster. Essentially the same age, but with larger uncertainty, is obtained 
using other samples, as shown in Figure \ref{f:6819}. 
Although somewhat on the young side, our age for NGC\,6819 is 
in good agreement with a number of other age determinations based on  
colour-magnitude diagram fitting \citep[e.g.,][]{rv98,krf01,yang13}, seismic 
masses \citep{basu11}, white dwarf cooling sequence \citep{bsa15} and 
eclipsing binaries \citep{sand13,jef13}. Values in the literature range from 
$2.0$ to $2.7$\,Gyr. Part of these differences depends on the models used in 
each study, as well as on the reddening and metallicity adopted for the  
cluster. We remark though the nearly perfect agreement with the age of
$2.25\pm0.20$\,Gyr from the white dwarf cooling sequence and $2.25\pm0.3$\,Gyr 
from the main-sequence turn-off match when using the same BaSTI models
\citep{bsa15}.

\begin{figure}
\begin{center}
\includegraphics[scale=0.27]{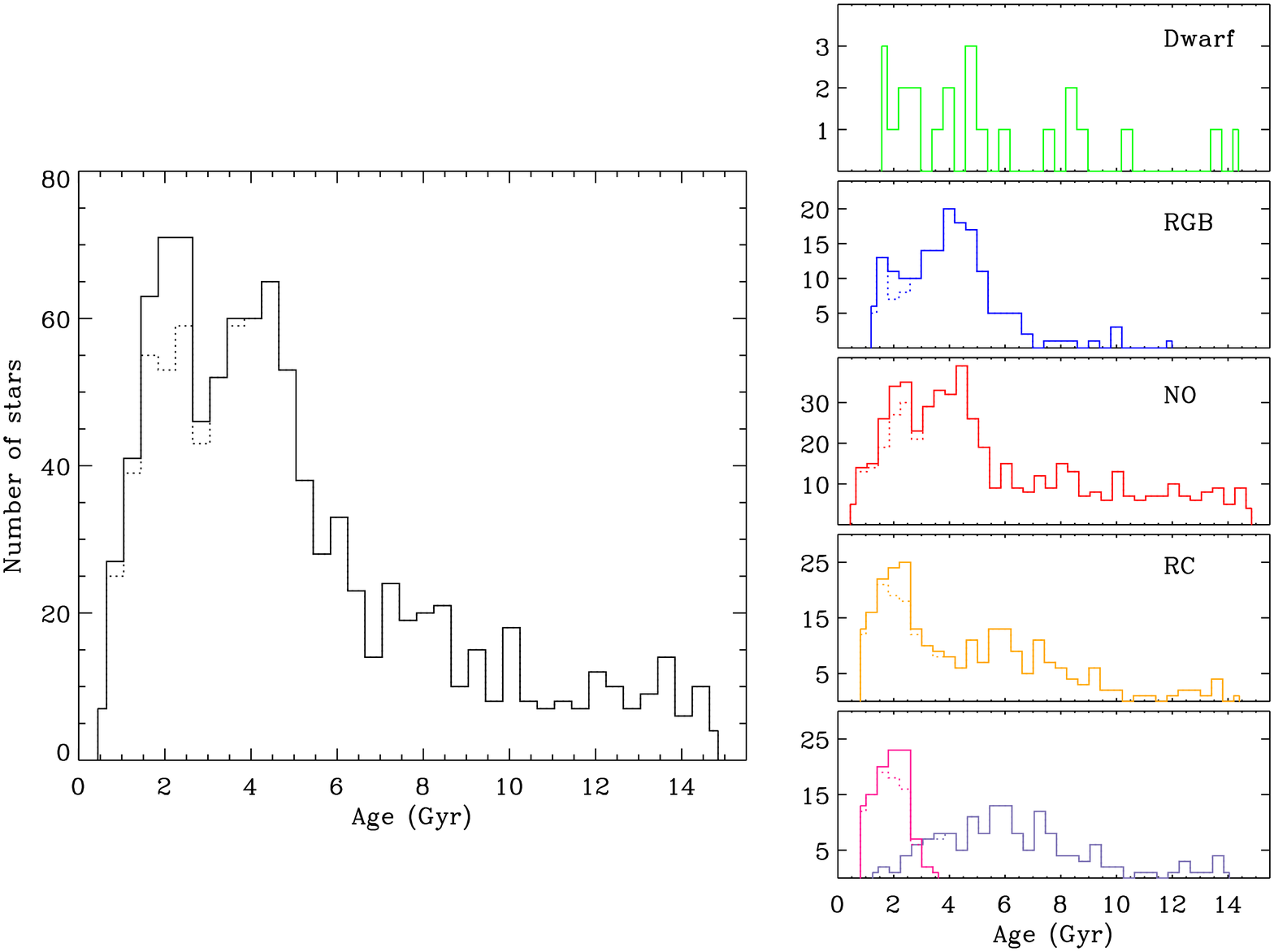}
\caption{Age distribution of the entire asteroseismic sample ({\it central
panel}), and when splitting stars according to seismic classification ({\it
right-hand side panels}). The lowest right panel is the age distribution 
of stars having certain ``RC'' classification and sorted into primary (blue) 
and secondary (pink) clump according to their $\logg$ (see description in the 
text). Dotted lines are same distributions once members of the cluster 
NGC\,6819 are excluded.}
\label{f:fs}
\end{center}
\end{figure}

Moving to the entire asteroseismic sample, Figure \ref{f:fs} shows the age
distribution of all stars, which peaks between 2 and 4\,Gyr. While this
distribution is not a proof of the reliability of the ages in itself, the
ability to single out a population of ``known'' ages is. Such a population is
provided by secondary clump stars, which are bound to be young 
\citep[$\lesssim 2$~Gyr, see e.g.][and also Section \ref{sec:rc2} for the use
of secondary clump stars as standard age candles]{girardi99}.

The distribution in Figure \ref{f:fs} varies quite considerably when split
according to seismic classification, shown on the right-hand panels. While
the ``RGB'' sample clusters at young ages, ``RC'' stars peak around 2\,Gyr
with a tail at older ages. The age distribution of stars without seismic
classification (which includes clump, upper and lower red giant branch, and
asymptotic giant branch stars) is a mix of the two previous distributions
characterized by a somewhat thicker tail at old ages.
The lowest right-hand panel in Figure \ref{f:fs} shows ages for ``RC''
stars, sorted into the primary or secondary clump phase. It is
important to stress that the distinction between primary and secondary clump
stars is done here with a (rather arbitrary) cut at $\logg=2.5$. Thus, there is
a certain level of contamination between the two phases, which surely broads the
age distribution of plausibly secondary clump stars.
In addition there is also contamination from members of NGC\,6819 which peaks
around 2\,Gyr. Once the seismic members of the cluster are excluded, the
typical age of secondary clump stars
shifts to younger values, in accordance with expectations \citep[e.g.,][]
{girardi99}, providing futher confidence on our asteroseismic ages.
Should the same figure be done using ages derived with mass-loss, the overall 
distributions would remain quite similar, but the tail at older ages would be 
reduced, in particular for ``RC'' stars.

The above comparisons tell us that despite the various uncertainties 
associated with age determinations, our results are meaningful. On an absolute 
scale, the age we derive for the open cluster NGC\,6819 is in agreement with 
the values reported in literature using a number of different methods. This 
holds at the metallicity of this cluster, which nevertheless is representative 
of the typical metallicity of most stars in the \kepler field. On a 
differential scale, once ``RC'' stars are identified as primary or secondary,
they show different age distributions. Despite our rough $\logg$ criterium
might partly blur this difference, the ability to recover the presence of
young secondary clump stars gives us further trust on our ages. 

\section{Statistical properties of the asteroseismic sample}\label{sec:stat}
\begin{figure*}
\begin{center}
\includegraphics[scale=0.17]{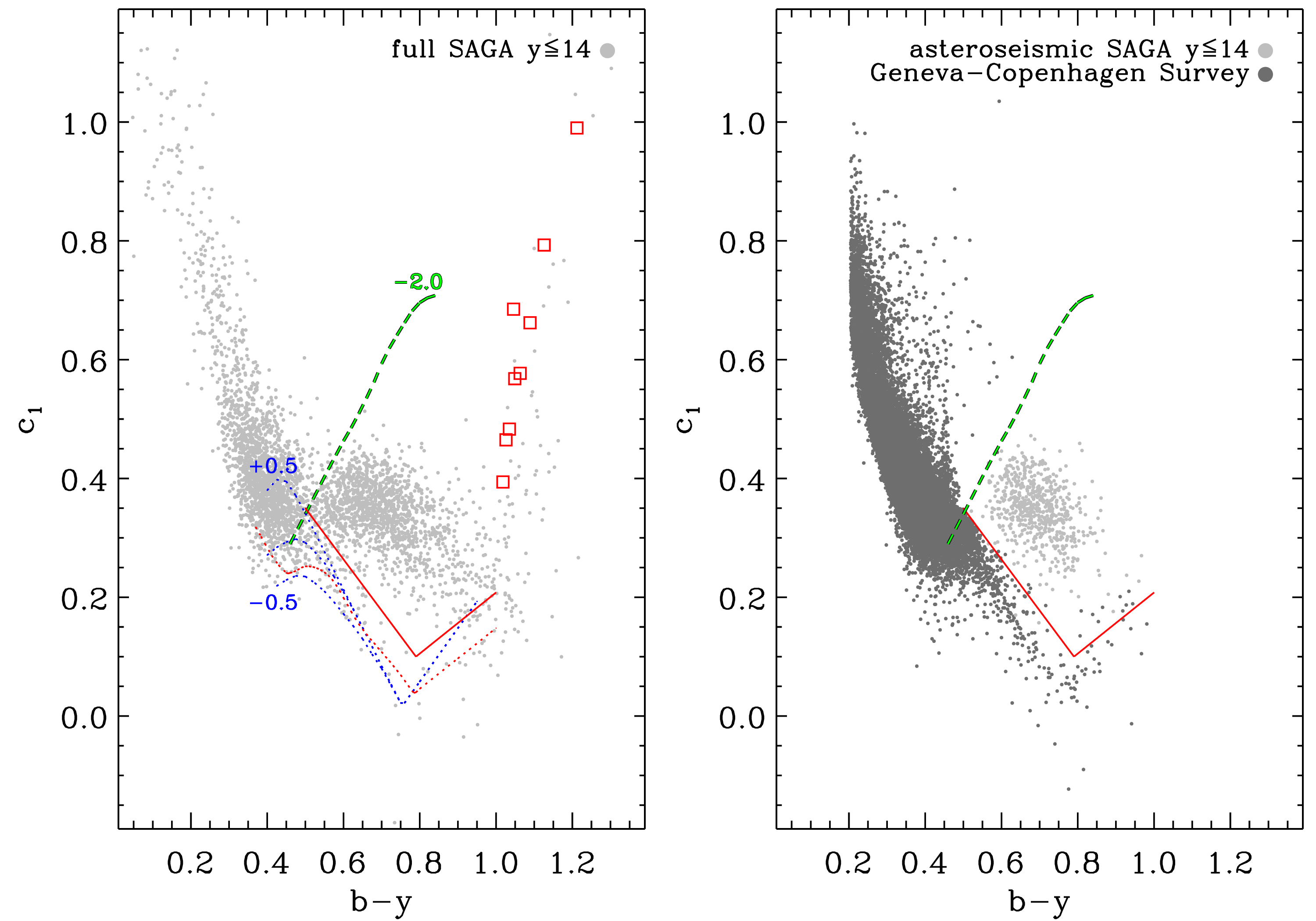}
\caption{{\it Left panel:} $b-y$ vs.~$c_1$ plane for the entire SAGA 
photometric catalog with $y \le 14$ and Pflg={\tt 0}. Dotted red line is the 
main-sequence fiducial of \citealt{olsen84}, while the continuous red line is
the $+0.06$~mag shift we use to separate dwarfs from giants. Open squares are 
cool M giants, also from \citealt{olsen84}. For reference, the metallicity
dependent dwarf sequences of \citealt{arnadottir10} are also shown (in blue,
for 
$-0.5 \leq \feh \leq 0.5$ as indicated), as well as the metal-poor
($\feh=-2.0$) red giant branch sequence of \citealt{att94}. {\it Right panel:}
dark gray circles identify all dwarf/subgiant stars in the Geneva-Copenhangen
Survey, most of the late-type ones being succesfully delimited by our shifted
fiducial (in red). Pale gray circles are all asteroseismic giants having good
photometric and metallicity flags and $y \le 14$.}
\label{f:byc1}
\end{center}
\end{figure*}

In order to use our sample for investigating age and metallicity gradients in 
the Galactic disc, we need to know how stars with different properties are 
preferentially, or not, observed by the \kepler satellite. In other words, we 
need to know the \kepler selection function. 

The selection criteria of the satellite were designed to optimize the 
scientific yield of the mission with regard to the detection of Earth-size 
planets in the habitable zone of cool main-sequence stars \citep{bata}. Even
so, deriving the 
selection function for exoplanetary studies is far from trival 
\citep[][]{phm13,ccb14}. For the sake of asteroseismic studies, entries in the 
KASC sample of giants (c.f. with Section \ref{subsec:as}) are based on a 
number of heterogeneous criteria 
\citep[][]{huber10,pin14}. Fortunately, the full \stromgren catalog offers a 
way of assessing whether seismic giants with particular stellar properties 
are more likely (or not) to be observed by the \kepler satellite. 

\subsection{Constraining the \kepler selection function}

Stellar oscillations cover a large range of timescales; for solar-like 
oscillations 
--as we are interested here-- these range from a few minutes in dwarfs 
\citep[cf e.g., with 5 minutes in the Sun,][]{lns62} to several days or more 
for the most luminous red giants \citep[e.g.,][]{derid09,dbs09}. 
The \kepler satellite has two observing modes: short-cadence (one minute), for 
dwarfs and subgiants \citep[a little over 500 objects with measured 
oscillations in the \kepler field, see][]{chap11,chap14} and long-cadence 
(thirty minutes) well suited for detecting oscillations in red giants.

With the exception of a few hundreds of dwarfs, most of the stars for which 
\kepler measured oscillations are giants. In order to assess how well these 
stars represent the underlying stellar population of giants, we use the full 
photometric catalog to build an unbiased sample of giants with well-defined 
magnitude and colour cuts. This task is facilitated both by the relatively 
bright magnitude limit we are probing, meaning that within a colour range most 
late-type stars are indeed giants, as well as by the fact that \stromgren 
colours offer a very powerful way to discriminate between cool dwarfs and
giants. We use the $(b-y)$ vs.~$c_1$ plane, which due to its sensitivity to
$\teff$ and $\logg$ (in the relevant regions), can be regarded as the
observational counterpart of an H-R diagram \citep[e.g.,][]{c75,olsen84,sbm04}. 
Working in the $(b-y)$ vs.~$c_1$ plane also avoids any metallicity selection 
on our sample. In fact, as we discuss below, we build our unbiased sample 
using cuts in $b-y$ colour, whereas metallicity acts primarily in a direction 
perpendicular to this index, by broadening the distribution of stars along
$c_1$. 

Figure \ref{f:byc1} shows the $(b-y)$ vs.~$c_1$ plane for the full photometric 
SAGA sample when restricted to $y \le 14$ mag, approximately 
the magnitude limit of the asteroseismic sample (a more precise magnitude 
limit will be derived in the next Section). This diagram is uncorrected for 
reddening, which is relatively low in the SAGA Galactic stripe studied 
here\footnote{Further, $E(b-y) \sim 0.75 E(B-V)$ and $E(c_1) \sim 0.2E(b-y)$, 
  reddening thus having limited impact on these indices \citep{cb70}.}. In
particular, in the following we focus on giants, all located across the same
stripe and having similar colours and magnitudes, meaning that 
reddening affects both the asteroseismic and the photometric sample in the
same way.

In the left-hand panel of Figure \ref{f:byc1}, gray dots nicely map the 
sequence of hot and turn-off stars for $b-y \lesssim 0.5$, whereas 
the giant sequence starts at redder colours, then upturning into the M 
supergiants at $b-y \gtrsim 1.0$. At the beginning of the giant sequence there 
is also an under-density of stars, consequence of the quick timescales in
this phase and mass regime (Hertzsprung gap). Below the giants is 
the dwarf sequence, here poorly populated because of our bright magnitude 
limit. To exclude late-type dwarfs from the full photometric catalog, we start 
from the \cite{olsen84} fiducial (dotted red line), which is representative of 
solar metallicity dwarfs. Since metallicity spreads the dwarf sequence, we 
shift Olsen's fiducial by increasing its $c_1$ by $+0.06$~mag, as shown in 
Figure \ref{f:byc1} (continuous red line). For this shifted fiducial, the 
linear 
shape at $b-y\simeq 0.55$ 
is more appropriate to exclude metal-rich dwarfs \cite[c.f. with][]
{arnadottir10}, and it fits well the upper locus of dwarfs in the GCS (shown 
in the right-hand panel, dark-gray dots). For $b-y >0.5$, our shifted fiducial 
extracts $687$ of the $704$ dwarfs in the GCS, thus proving successful to
single out dwarfs from giants ($>97$ per cent). Also shown for comparison is an 
empirical sequence for metal-poor giants (green dashed line, from 
\citealt{att94}). 
Indeed, almost all of the targets with $b-y \gtrsim 0.5$, including the
asteroseismic giants, lie on the right-hand side of this
metal-poor sequence (as expected, given the typical metallicites encountered 
in the disc) thus indicating that an unbiased selection of giants is possible 
in the $b-y$ vs.~$c_1$ plane. 

To summarize, any unbiased, magnitude-complete sample of giants used in this 
investigation will be built by selecting giants from the full photometric  
catalog in the $b-y$ vs.~$c_1$ plane, with the colour and magnitude cuts we 
will derive further below. Aside from being 
shown for comparison purposes, the giant metal-poor sequence discussed above 
is not used in our selection, while the shifted Olsen's fiducial derived above
is employed to avoid contamination from dwarfs. We remark again that at the 
bright magnitudes studied here, contamination from dwarfs is expected to be 
minimal, most stars with late-type colours being in fact giants.

\begin{figure}
\begin{center}
\includegraphics[scale=0.08]{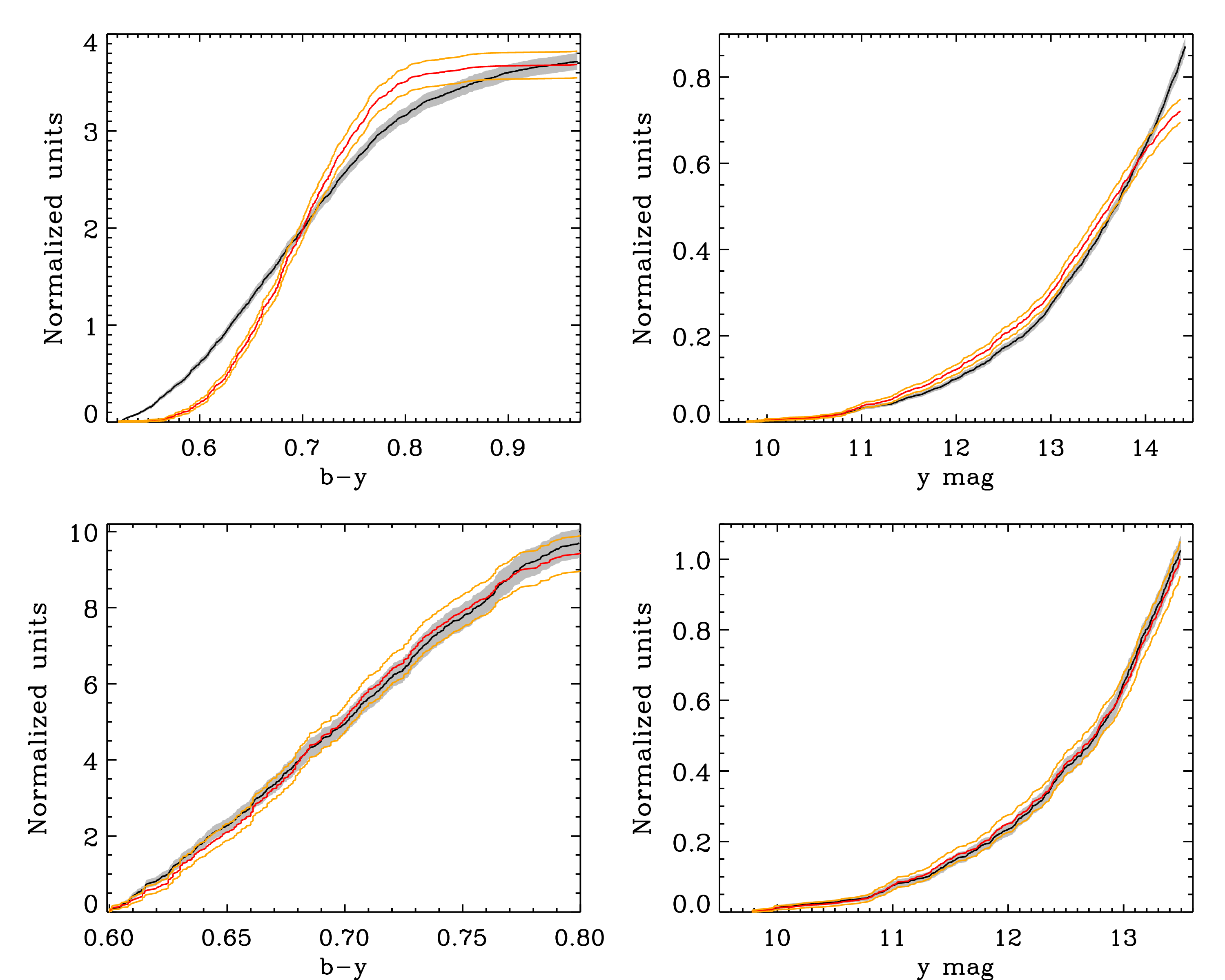}
\caption{{\it Top panels:} cumulative distribution in $b-y$ and $y$ for the 
the unbiased photometric sample of giants (in black, with gray shaded area 
indicating $1\sigma$ Poisson errors) and the uncut asteroseismic sample of
giants 
(in red, with orange line indicating $1\sigma$ Poisson errors) having the same 
colour and magnitude limits ($0.52 \le b-y \le 0.97$ and $y \le 14.4$). All 
curves are normalized to equal area. {\it Low panels:} same as above, but 
restricting both the asteroseismic and the unbiased photometric sample of
giants to $0.6 \le b-y \le 0.8$ and $y \le 13.5$.}
\label{f:cumplot}
\end{center}
\end{figure}

To derive the appropriate magnitude and colour cuts, we first explore how the 
asteroseismic sample of giants compares with the unbiased sample of giants 
built with the same magnitude limit ($y=14.4$) and colour range 
($0.52 \le b-y \le 0.97$) comprising the asteroseismic one. Should the latter
be 
representative of the underlying population of giants within the same colour 
and magnitude limits, we would expect the relative contribution of giants 
at each colour and magnitude be the same for both the asteroseismic and the 
unbiased sample. This comparison is performed 
in the two upper panels of Figure \ref{f:cumplot}, for the unbiased 
photometric sample 
of giants (black line, with gray dashed area representing $1\sigma$ Poisson 
errors) and the asteroseismic giants (red line, with orange contour lines 
representing $1\sigma$ Poisson errors). Since the total number of stars is 
different in the two samples, all curves are normalized to equal area. 
It is clear from both panels that the asteroseismic and the unbiased sample of 
giants have different properties: in fact the asteroseismic sample has 
considerably fewer stars towards the bluest (hottest) and reddest 
(coolest) colours (effective temperatures). In addition, the asteroseismic 
sample begins to lose stars at the faintest magnitudes. 

Although the selection of seismic targets by {\it Kepler} was 
heterogeneous, and not intended for studying stellar populations 
in the Galaxy, the observed selection effects are understandable: stars with 
bluer colours (hotter $\teff$) are at the base of the red giant branch, where
stars oscillate with intrinsically smaller amplitudes, and the {\it Kepler}
long-cadence mode (thirty minutes) also becomes insufficient to sample the
shorter oscillation periods of these stars. 
Conversely, on the red side, moving along the red giant branch towards 
cooler $\teff$ and brighter intrinsic luminosities, the timescale of 
oscillations increases, until the characteristic frequency separation can no
longer be resolved robustly with the length of our \kepler observations (up to
Quarter 15, i.e.~typically well over 3 years). 

In order to have an unbiased asteroseismic sample, we must avoid the 
incompleteness towards the bluest and reddest colours as well as at the
faintest magnitudes. We explore different cuts in $b-y$ and 
$y$, finding that for $0.6 \le b-y \le 0.8$ and $y \le 13.5$ the asteroseismic 
sample is representative of the underlying population of giants in the same 
colour and magnitude range. To this purpose, we use the Kolmogorov-Smirnov 
statistic on the colour and magnitude distributions: the significance levels 
between the asteroseismic and the unbiased 
photometric sample in $b-y$ and $y$ pass from $\sim 10^{-6}$ and $\sim 10^{-11}$ 
to about $67$ per cent and $98$ per cent respectively, when we use the cuts
listed above. 
This implies that the null assumption that the two samples are drawn 
from the same population can not be rejected to a very high significance. 
Equally high levels of significance are obtained for the other \stromgren 
indices $m_1$ (73 per cent) and $c_1$ (99 per cent), as well as when the two
samples are compared as function of Galactic latitude $b$ (94 per cent). 
For all the above parameters, significance levels of $\simeq 20$ to $80$
per cent  are also obtained using different statistical indicators such as the 
Wilcoxon Rank-Sum test, and the F-statistic. We remark however that the 
unbiased photometric sample (641 targets) also includes all the seismic 
targets (408) within the same magnitude and colour limits. To relax this 
condition, we bootstrap resample the datasets $10,000$ times and find that 
significance levels for all of the above tests vary between $30-60$ per cent
when bootstrapping either of the two samples, to $20-40$ per cent when
bootstrapping both. 
Since for all these tests significance levels below 5 per cent are generally
used to discriminate whether two samples originate from different populations,
we thus conclude that the asteroseismic sample is representative of the
underlying population of giants to a very high confidence level.
\begin{figure}
\begin{center}
\includegraphics[scale=0.46]{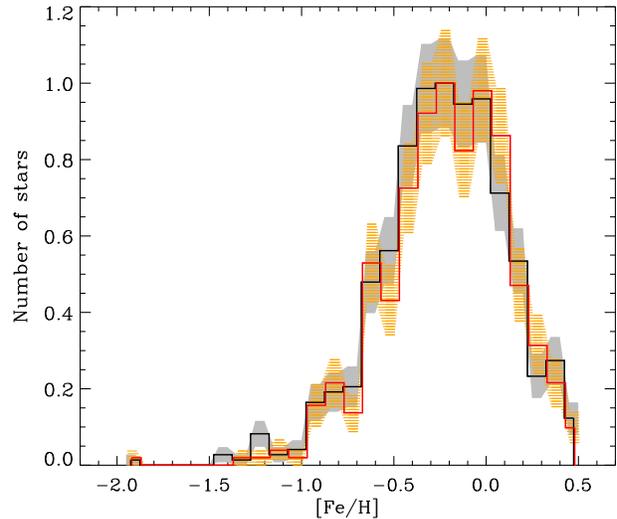}
\caption{Normalized metallicity distribution for the unbiased photometric 
  sample of giants (black line, with shaded gray area indicating $1\sigma$
  Poisson errors) and the asteroseismic sample (red line, with shaded orange
  area indicating $1\sigma$ Poisson errors) in the same magnitude and colour
  range $0.6 \le b-y \le 0.8$ and $y \le 13.5$. Only stars with good
  photometric and metallicity flags are used.}
\label{f:mdf}
\end{center}
\end{figure}

Our photometry is significantly affected by binarity only in the case of near 
equal luminosity companions (or equal mass, dealing with giants at the same 
evolutionary stage). These binaries imprint an easily recognizable signature
in the seismic frequency spectrum and are very rare (5 such cases in the 
full SAGA asteroseismic catalog, see discussion in Paper~I). When restricting 
to the unbiased asteroseismic catalog three such cases survive, implying an
occurrence of near equal-mass binaries of $0.7 \pm 0.4$ per cent. We exclude
these binaries from the analysis. Although we cannot exclude such binaries
from the photometric sample, we expect the same fraction as in the
asteroseismic catalog. All of the above statistical tests remain 
unchanged whether the asteroseismic sample with (411) or without binaries (408 
targets) is benchmarked against the unbiased photometric catalog of giants, 
suggesting that indeed they have a negligible effect on our results.

From the above comparisons, we have already concluded that the asteroseismic
sample 
of giants is representative of the underlying populations of giants in both 
colour and magnitude distribution. We expect this to be true for all other 
properties we are interested in as well. Whilst this is impossible to verify 
for masses and ages, Str\"omgren photometry offers a convenient way of 
checking this in metallicity space. 

Before deriving photometric metallicities we must correct for reddening also 
the unbiased photometric sample (in fact, metallicities for the asteroseismic
sample were derived after correcting for reddening). Interstellar extinction
is rather low and well constrained in the magnitude range of our targets; we
fit the $E(B-V)$ values of the asteroseismic sample as an exponential function
of Galactic latitude; this functional form reflects the exponential disc used
to model the spatial distribution of dust in the reddening map adopted for
the asteroseismic sample (see Paper~I for a discussion). The fit has a scatter
$\sigma=0.015$~mag, 
which is well within the uncertainties at which we are able to estimate 
reddening. More importantly, despite this fit is based upon the asteroseismic 
sample (the one we want to estimate biases upon), it also reproduces 
(within the above scatter) the values of $E(B-V)$ obtained from 2MASS, using 
an independent sample of several thousand stars (see details in Paper 
I). Such a fit 
obviously misses any three-dimensional information on the distribution of dust, 
but the purpose here is to derive a good description of reddening for the 
population as a whole, in the range of magnitudes, colours, and Galactic 
coordinates covered in the present study. 

After correcting for reddening, we apply the same giant metallicity
calibration used for the asteroseismic sample (Paper I) to the photometric
unbiased 
sample of giants, and compare the two (Figure \ref{f:mdf}). In both cases we 
only use stars with good photometric and metallicity flags 
(i.e.~when the calibration is applied within its range of validity). 
We run the same statistical tests discussed above also for the distributions 
in metallicity, and significance levels varies between $15$ and $50$ per cent  
depending on the test and/or whether bootstrap resampling is implemented or 
not. Based on the above tests, we can thus conclude that for $y \le 13.5$ and 
$0.6 \le b-y \le 0.8$ the $\feh$ distribution of the asteroseismic sample
represents that of the giants in the field within the same colour and magnitude
ranges.

\begin{figure}
\begin{center}
\includegraphics[scale=0.12]{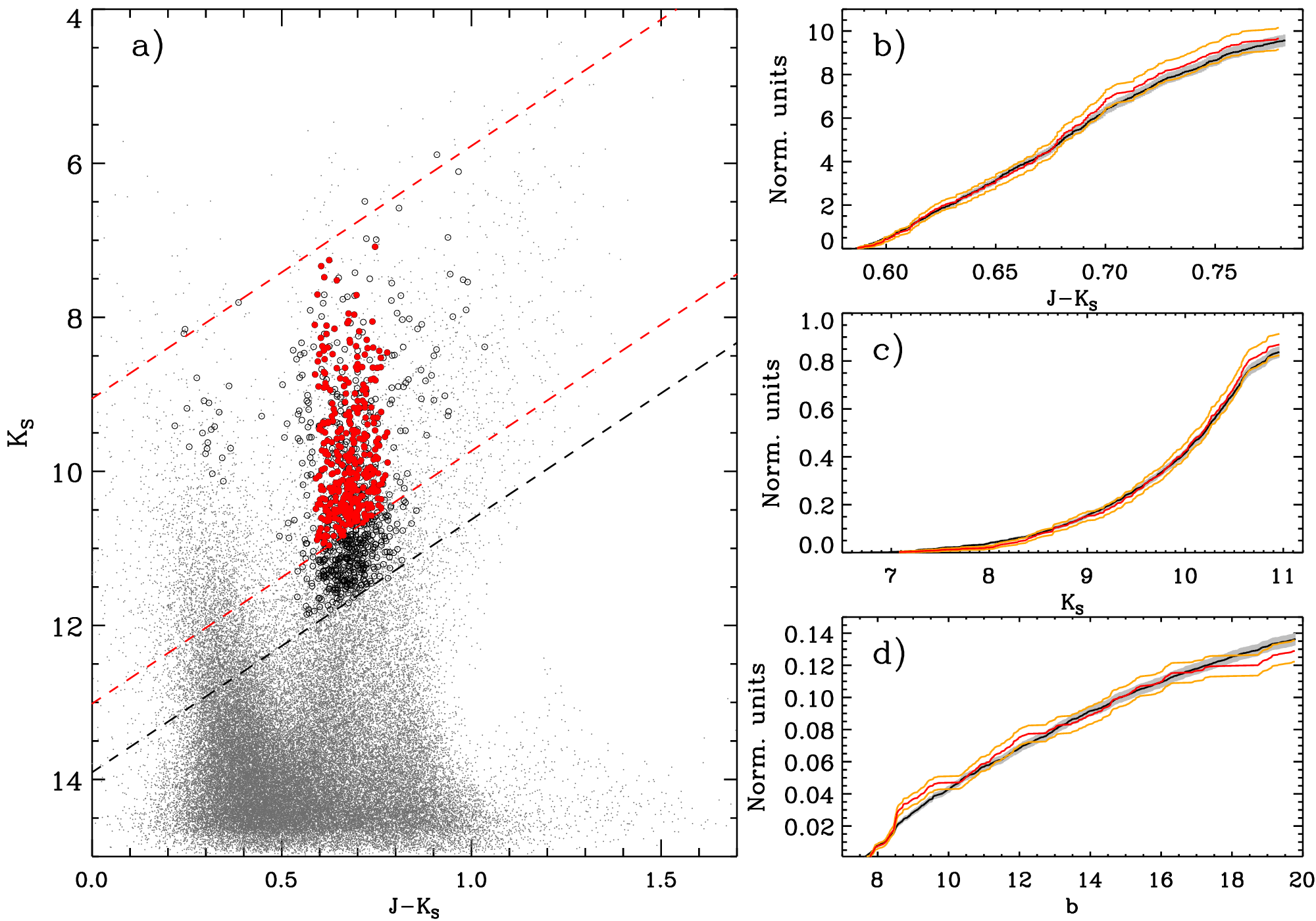}
\caption{{\it Left panel:} 2MASS $K_S$ vs.~$J-K_S$ diagram (gray dots) for 
  stars in approximately the same Galactic stripe of the asteroseismic sample
  (black open circles). Filled red circles identify stars belonging to the
  unbiased asteroseismic sample built as described in the text, i.e.~having
  good photometric and metallicity flags, and with $0.586 \le J-K_S \le 0.782$
  and $K_S \le 13.01 - 3.28\,(J-K_S)$. {\it Right panels:} cumulative
  distributions between the unbiased 2MASS photometic sample and seismic
  giants with same colour and magnitude cuts (colour code same as of Figure
  \ref{f:cumplot}).}
\label{f:2MASS}
\end{center}
\end{figure}

Although we have already constrained the \kepler selection function using our
\stromgren photometry, we also explore whether 2MASS photometry offers an
alternative approach of assessing it, for the sake of other dataset where
\stromgren is not available \citep[e.g., such as APOKASC,][]{pin14}.
In Figure \ref{f:2MASS}a), dark-gray dots show the $K_S$ vs.~$J-K_{S}$
colour-magnitude diagram for stars approximately in the same stripe of the 
asteroseismic sample ($73.4^{\circ} \le l \le 74.4^{\circ}$ and 
$ 7.6^{\circ} \le b \le 19.8^{\circ}$). In this plot, three main features 
are obvious: the overdensity of stars around $J-K_S \simeq 0.35$, 
which corresponds to main-sequence and turn-off stars; the overdensity at 
$J-K_S \simeq 0.65$ comprising primarily giants, and the blob at 
$J-K_S \simeq 0.85$ and faint magnitudes ($K_S \gtrsim 13$), mostly comprising 
cool dwarfs. 
Again, at bright magnitudes most late-type stars are giants. Overplotted with 
circles is the entire asteroseismic sample, including both dwarfs and giants 
independently of their Mflg and Pflg flags. The $K_S$ magnitude limit of 
{\it Kepler} is clearly a function of spectral type, or $J-K_S$ colour. Using 
seismic giants only, we derive the following relation between 
\stromgren and 2MASS magnitudes: $K_S=0.99\,y - 3.28 (J-K_S) -0.35$, with a 
scatter $\sigma_{K_{S}}=0.09$ mag. The inclined black-dashed line in Figure 
\ref{f:2MASS}a) corresponds to a constant $y=14.4$, which, as we previously 
saw, is roughly the limit of the faintest stars selected to measure
oscillations in {\it Kepler}.
At bright magnitudes, we introduce a similar cut, corresponding to $y=9.5$ 
(upper red-dashed line) which is approximately the saturation limit of the INT. 
This bright cut removes only a handful of stars and it is of limited 
importance. If we now compare the asteroseismic sample of giants (i.e.~all
giants stars with 
good Mflg and Pflg flags) with the entire 2MASS sample within the same 
magnitude limit ($9.5 \le \frac{K_S+0.35+3.28(J-K_S)}{0.99} \le 14.4$) and 
colour range ($0.494 \le J-K_S \le 0.951$), the hypothesis that two samples are 
drawn from the same population is rejected. The same is still the case if we
use a constant magnitude cut $6 \le K_S \le 12$ (such to encompass our sample,
see Fig.\,\ref{f:2MASS}a) instead of the colour dependent one done above. 

From the \stromgren analysis we already know the magnitude and colour range 
where the asteroseismic targets are expected, on average, to unbiasedly sample
the underlying population of giants. Thus, we can see how these limits
convert in the 2MASS system. Using all seismic targets we
derive the 
following relation $J-K_S=0.977\,(b-y)$ with $\sigma_{J-K_{S}}=0.04$, which 
converts $0.6 \le b-y \le 0.8$ into $0.586 \le J-K_S \le 0.782$. We also apply 
a colour dependent $K_S$ magnitude cut corresponding to $9.5 \le y \le 13.5$ 
(filled red circles and red-dashed lines in Figure \ref{f:2MASS}a). In this 
case, 
the seismic and 2MASS samples are drawn from the same population to 
statistically significant levels in colour, magnitude and Galactic latitude 
distribution, as qualitatively shown on the right-hand side panels of
Figure \ref{f:2MASS}.

\section{Vertical Mass and Age Gradients in the Milky Way disc}\label{sec:grad}

\begin{figure*}
\begin{center}
\includegraphics[scale=0.25]{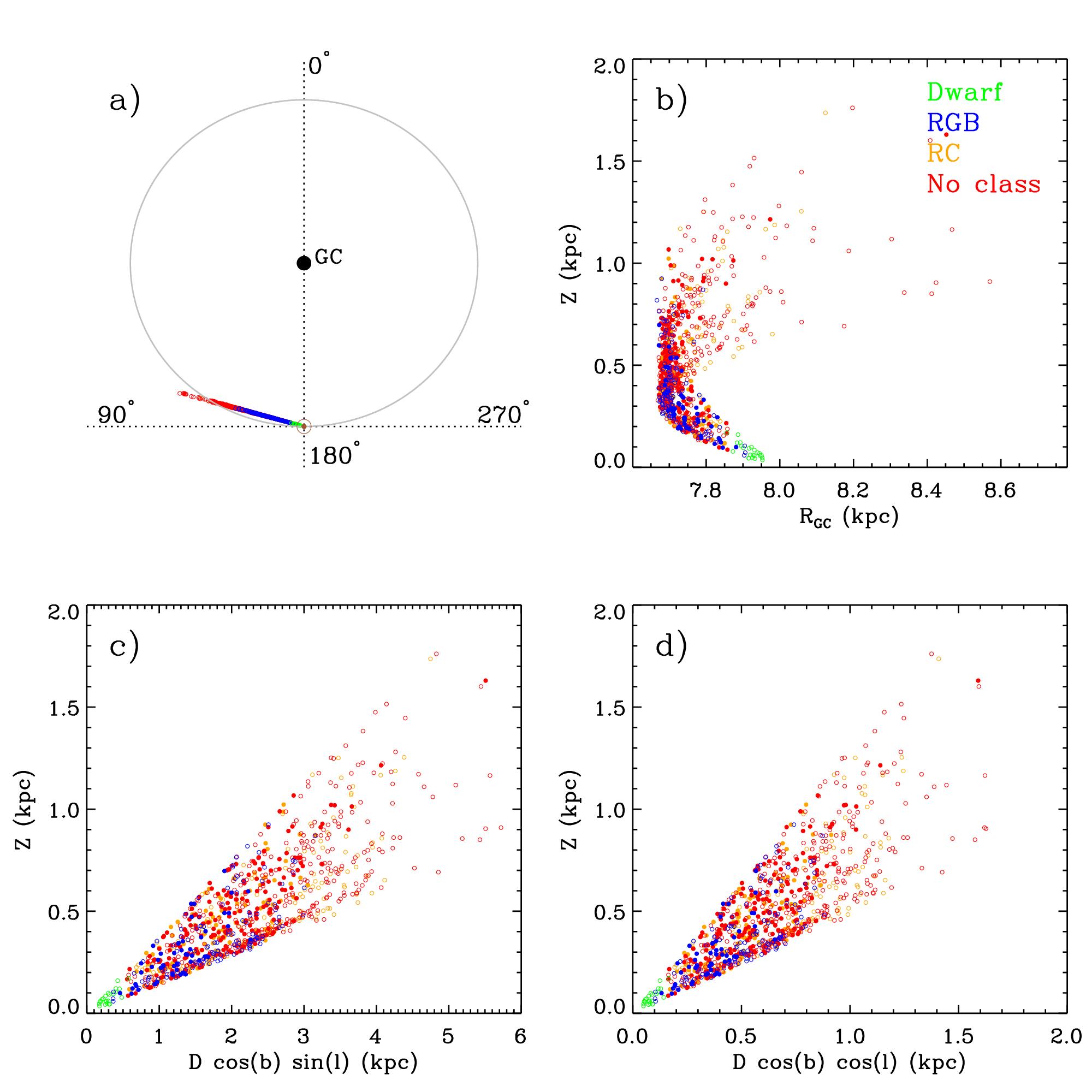}
\caption{Location of the SAGA targets in Galactic coordinates. Stars with
  different colours have different seismic evolutionary classification
  ``Dwarf'', ``RGB'', ``RC'' and ``NO'' as indicated. Filled circles identify 
the 373 stars which satisfy the constraints described in Section \ref{sec:grad}.
{\it Panel} a): target distribution over the Galactic plane, where the distance 
of each seismic target from the Sun ($D$) is projected along the line of sight 
$D\,\cos(b)$ having direction $l\simeq 74^{\circ}$ and Galactic latitude $b$. 
The distance between the Galactic Centre (GC) and the location of the Sun 
($\odot$) is marked by the gray circle. Galactic longitudes ($l$) 
at four different angles are indicated. {\it Panel} b): same as above, but as 
function of Galactic height $\textrm{Z} = D\,\sin(b)$ and Galactocentric Radius 
($\rm{R_{GC}}$, computed assuming a solar distance of $8$~kpc from the 
Galactic Centre). {\it Panel} c) and d): Z distribution of targets across 
two orthogonal directions. The multiple beams structure in panels 
b) to d) arises from the projection of the gaps in the CCD modules
on {\it Kepler}.}
\label{f:xyzr}
\end{center}
\end{figure*}

\begin{figure*}
\begin{center}
\includegraphics[scale=0.25]{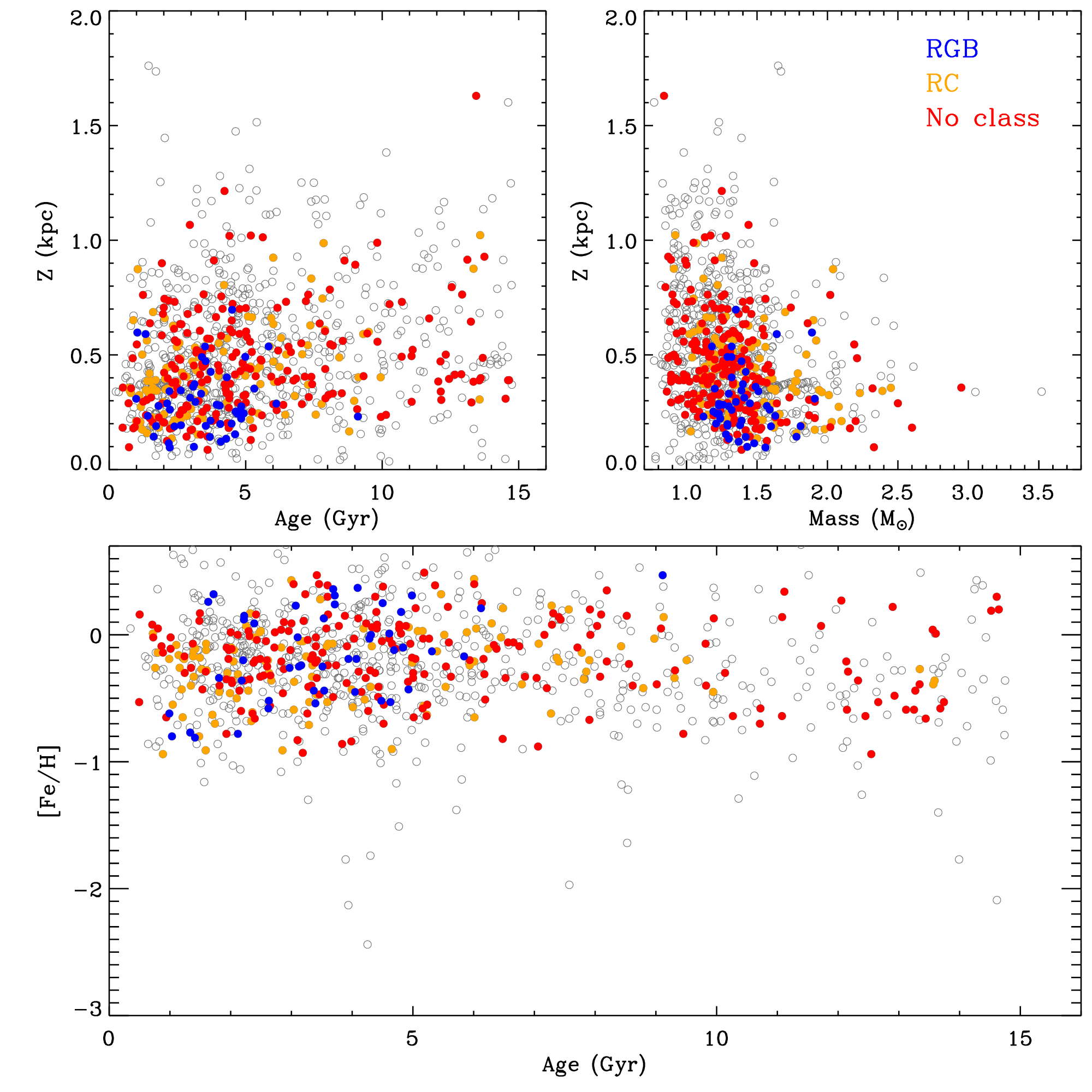}
\caption{Raw age and mass gradients ({\it top panels}) and age-metallicity
  relation ({\it lower panel}) before 
correcting for completness and target selection. Open circles are all giants 
in SAGA, while filled circles (colour coded according to their seismic 
classification) are stars satisfying the seismic target selection discussed in 
Section \ref{sec:ss}.}
\label{f:wholehog}
\end{center}
\end{figure*}

The \kepler field encompasses stars located in the direction of the
Orion arm, edging toward the Perseus, and rising above the Galactic plane.
The stripe observed so far by SAGA has Galactic longitude $l\simeq 74^{\circ}$
and covers latitude 
$8^{\circ} \lesssim b \lesssim 20^{\circ}$. Its location in the Galactic 
context is shown in Figure \ref{f:xyzr}; it can be immediately appreciated from
panel a) and b) that the geometry of the SAGA survey 
allows us to probe distances of several kpc from the Sun at nearly the same 
Galactocentric radius, thus minimizing radial variations and greatly 
simplifying studies of the vertical structure of the Milky Way disc. At the 
same time, SAGA spans a vertical distance Z (altitude or height, hereafter) of 
about 
$1.5$~kpc, probing the transition between the thin and the thick disc, 
which have scale-heights of $\simeq 0.3$ and $\simeq 1$~kpc, respectively 
\citep[e.g.,][]{juric08}. In Figure \ref{f:wholehog} we show the raw 
dependence of stellar ages and masses with Galactic height, as well as the raw
age-metallicity relation. For the purpose of Galactic studies, these plots can 
not be taken at face value, and must first be corrected for target 
selection effects (stemming from the colour and magnitude cuts derived in the 
previous Section), as well as to account for the fact that in the most
  general case, the ages of red giants might not be representative of those of
  an underlying stellar population. These steps are described further below. 

\subsection{Sample selection}\label{sec:ss}

To estimate gradients, we limit our sample to $0.6 \le b-y \le 0.8$ and 
$y \le 13.5$ such that it reflects the distribution of the underlying 
sample of giants (Section \ref{sec:stat}). We remove stars labelled as 
binaries and those flagged as having poor photometry and/or metallicity 
estimates (see Section \ref{sec:SAGA}). The latter requirement automatically 
excludes stars with $\feh \geqslant 0.5$, but we also limit the metallicity
range to $\feh > -1.0$, to remove any halo object, which would contaminate
our study of disc gradients\footnote{Note that our colour cut alone already
  removes many of the metal-poor objects.}. By excluding metal-poor stars we
also avoid problems related to the potential inaccuracy of seismic scaling
relations in this regime \citep{eej14}. Because we are interested in studying 
properties of the Galactic disc via field stars, we also exclude members of the 
open cluster NGC\,6819, based on their seismic membership. Furthermore, we 
remove all targets classified as ``Dwarf'', obtaining a final sample of 373 
giants (i.e. with seismic evolutionary classification ``RGB'', ``RC'' or
``NO'', see 
Section \ref{subsec:as}). They cover heights from $\approx 0.1$ to $1.5$ kpc 
(Figures \ref{f:xyzr} and \ref{f:wholehog}). 
From the above sample of giants, we also extract a subsample of 48 seismically 
classified ``RGB'' stars with age uncertainties less than 30 per cent.
As discussed in Section \ref{sec:ages}, mass-loss can severely affect age
estimates of unclassified and clump stars, whereas ``RGB'' stars are essentially
immune to such uncertainty. These stars provide more robust ages, though at
the price of a greatly reduced sample size. We also refer to Paper I for the
uncertainties associated to masses and distances, which are 
of order 6 and 4 per cent, respectively. In the following we will determine
vertical gradients using both samples whenever possible: the 373 ``Giants'' and
the 48 best pedigreed ``RGB'' stars. The bulk of gravities for the ``Giants''
sample covers the range $2.0 < \logg < 3.5$, while for ``RGB'' stars covers 
$2.6 < \logg < 3.35$: we will use these values when modelling target selection
in Section \ref{sec:grad_bias}.

\subsection{Methodology and raw vertical gradients}\label{sec:grad_method} 

We adopt two methodologies to estimate the vertical gradient of age and
mass. First
{\it i)} we use a boxcar-smoothing technique described in
\citet{sch14}. 
Sorting the stars by height above the plane, we calculate the median age 
(mass) and altitude Z of a fraction of the sample at the lowest height. We then 
step through the sample in altitude, as we want to quantify the age (mass) 
variation with height above the plane. Each bin contains the same number of 
stars and overlaps by a small fraction with the previous bin. For the
``Giants'' sample, we explore the range between 18 and 30 stars per bin with
overlaps
ranging from 8 to 15. The ``RGB'' sample is much smaller and we explore the
range between 8 and 10 stars per bin with overlaps ranging from 2 to 4 stars. 
The binsizes and overlaps explored contain enough targets so that the overall
trend is not dominated by outliers, and the median points well reflect
the overall behaviour of the underlying sample. We then perform a
least-squares fit on these median points; the change in slope (i.e.~gradient)
due to different choices of binsize and overlap is typically below half the
uncertainty of the fit parameter itself.
We perform a Monte-Carlo to explore the sensitivity of the boxcar-smoothing on
the uncertainty of the input ages (masses), and add this uncertainty in
quadrature to those estimated above. We obtain the following raw age and mass
gradients for the ``Giants'' $3.9 \pm 1.1$ Gyr kpc$^{-1}$,
$-0.39 \pm 0.10\,\rm{M}_\odot$ kpc$^{-1}$. Similarly, for the ``RGB'' stars we
have $-0.1 \pm 3.3$ Gyr kpc$^{-1}$, $-0.09 \pm 0.35\,\rm{M}_\odot$ kpc$^{-1}$.

Our second estimate of the gradient {\it ii)} consists of a simple
least-squares fit to all of the stars that meet our criteria. Again our
uncertainties include those from the fitting coefficients and from a
Monte-Carlo. In this case we obtain for the ``Giants'' $4.1 \pm 0.9$ Gyr 
kpc$^{-1}$, $-0.43 \pm 0.08\,\rm{M}_\odot$ kpc$^{-1}$ and for the ``RGB'' stars 
$0 \pm 1.7$ Gyr kpc$^{-1}$, $0.03 \pm 0.20\,\rm{M}_\odot$ kpc$^{-1}$.

With both methods, the gradients for the ``RGB'' stars have considerably
larger uncertainties, which make them consistent with no slope and limit their
usability to derive meaningful conclusions. This is due to the small sample
size and scatter of the points. Because of this, the $\chi^2$ of the
``RGB'' fits have the same statistical significance whether we let the slope and
intercept be free, or we fix the latter on the ``Giants'' sample (roughly 3~Gyr
and $1.5\,\rm{M}_\odot$ on the plane). With this caveat in mind (i.e. fixing
the intercept), and including
in the error budget the uncertainty in the intercept derived from the
``Giants'', the raw ``RGB'' slopes become $3.6 \pm 1.7$ Gyr kpc$^{-1}$ and
$-0.43 \pm 0.17\,\rm{M}_\odot$ kpc$^{-1}$ for method {\it i)}, and
$1.4 \pm 1.2$ Gyr kpc$^{-1}$ and $-0.39 \pm 0.14\,\rm{M}_\odot$ kpc$^{-1}$
for method {\it ii)}. 

Technique {\it i)} and {\it ii)} have different strengths; as the
sample size is small, the least-squares fit takes full advantage of every star
available. However, the boxcar-smoothing technique avoids being skewed by any
outliers. Additionally, we can see how the uncertainties vary with respect to
height above the plane by examining the variation in each median point. 

We stress that both methods still need to be corrected for target selection
effects, i.e. the gradients above should not be quoted as the
values obtained for the Galactic disc. Also, the use of stellar masses
as proxy for stellar ages is applicable only to red giants. Thus, while it is
meaningful to derive a Galactic age gradient by assessing how well our sample
of red giants (with known selection function) will convey the age structure of
the larger underlying stellar population (done in the next Section), the
stellar mass gradient will reflect the mass structure of the underlying
population of red-giants only.
For red giants, the relation between mass and age is
$\textrm{Age}\propto M^{-\alpha}$, with $\alpha \simeq 2.5$. Thus, we expect
that the age gradient traced by red giants translates into a variety of masses
at the youngest ages, whereas low-mass (i.e.\,old) stars will be preferentially
found at higher altitudes. Indeed, this picture is consistent with Figure
\ref{f:wholehog}, which shows an L-shaped distribution of red-giants,
with low-mass stars extending from low to large heights and more massive stars
being preferentially close to the Galactic plane. Because of the aforementioned
power-law relationship between age and mass, one might wonder whether a linear
fit is appropriate
for quantifying the mass gradient shown by red giants. In fact, a change of
say $0.2 \rm{M}_\odot$ translates to a few 100~Myr in a $2 \rm{M}_\odot$ star,
but corresponds to several Gyr at solar mass. Here, our goal is not to provide
a value for the mass gradient --which given the above discussion would be of
limited utility-- but simply to use the masses of our red giants as a model
independent signature of the vertical age gradient. The above fits of the
mass gradient suffice for this purpose, and in the following discussion
we will focus only on the vertical age gradient.

\subsection{Correcting for target selection}\label{sec:grad_bias} 

In Section \ref{sec:stat} we have studied the {\it Kepler} selection function
to determine the colour and magnitude ranges in which the SAGA asteroseismic
sample reflects the properties of an underlying unbiased photometric sample of
red giants. However, to derive the Galactic age gradient we must assess how
target selection systematically affects our gradient estimates (i.e.~once a
clear selection function is defined, we must assess its effect). To avoid our
results being too depend on particular model assumptions, we use various
approaches to understand how our selection criteria and survey geometry will
bias our sample, and to what extent the ages of a population of red giants are
representative of the ages of a full stellar population.
\begin{figure*}
\begin{center}
\includegraphics[scale=0.7]{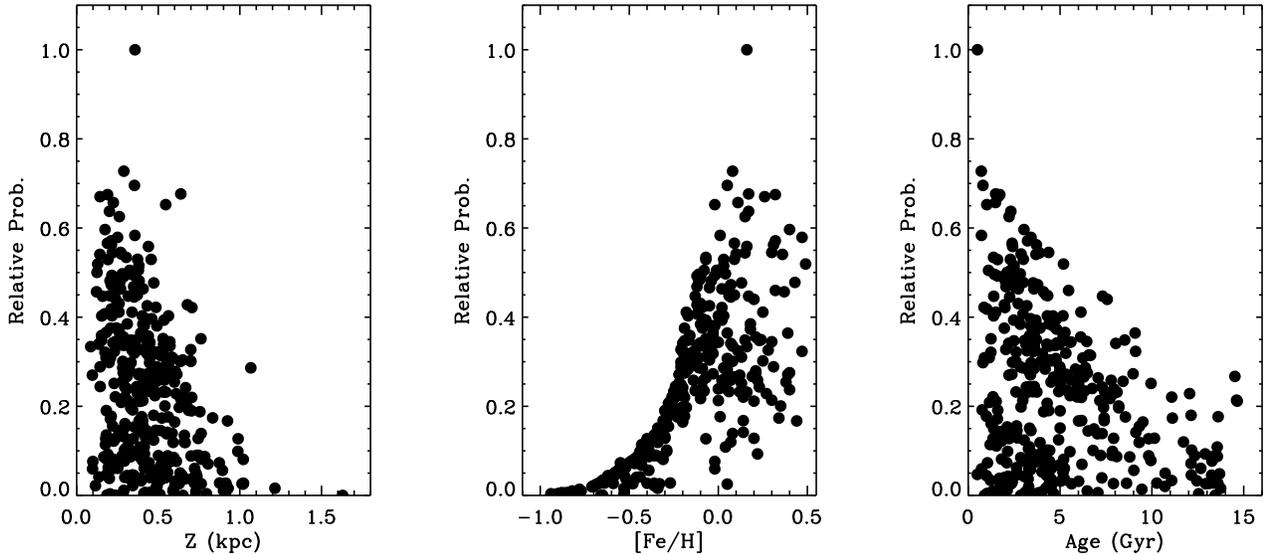}
\caption{Probability of a star passing the ``Giants'' target selection to be
  observed given its height, metallicity and age (see description in
  Section \ref{sec:nifty}). All probabilities are normalized to an arbitrary
  scale.}
\label{f:nifty}
\end{center}
\end{figure*}

\subsubsection{Target selection modelling}\label{sec:nifty}

We first want to examine the probability that a star with specific stellar
parameters will be observed given our
target selection criteria. We generate a data-cube in age, metallicity and
distance where each point in the age and metallicity plane is populated
according to a Salpeter IMF over the BaSTI isochrones. For each of these
populations we then assign apparent magnitudes by running over the distance
dimension in the cube. Thus, for each combination of age, metallicity and
distance we can define the probability of a star being observed by SAGA as the
ratio between how many stars populate that given point in the cube, and how
many pass our sample selection (i.e. our color, magnitude and gravity cuts,
see Section \ref{sec:ss}). This approach naturally accounts for the effects of
age and metallicity on the location of a star on the HR diagram. Via the IMF
it also accounts for the fact that stars of different masses have different
evolutionary timescales, and thus different likelihood of being age tracers of
a given population. This approach is the least model dependent, and provides
an elegant way to gauge the selection function.

Figure \ref{f:nifty} shows the probability of each star being observed given 
its height, metallicity and age. Our sample is biased against stars
at large distances (and thus altitudes), low metallicities, and old ages. 

We can then apply methodology {\it i)} and {\it ii)} described in Section
\ref{sec:grad_method}, where in the boxcar-smoothing/fitting procedure we
assign to each star a weight proportional to the inverse of its probability.
Stars with low probability will be given larger weight to compensate for the
fact that target selection is biased against them. Figure \ref{f:nifty}
indicates that probabilities are non linear functions of the input
parameters; for some targets the combination of age, metallicity and
distance results in a probability of zero, which then translates into an
unphysical weight. Observational errors are mainly responsible for scattering
stars into regions not allowed in the probability space. To cope with this
effect without setting an arbitrary threshold on the probability level, for
each target we compute the probability obtained by sampling the range of
values allowed by its age, metallicity and distance uncertainties with a
Monte-Carlo. While
this procedure barely changes the probabilities of targets very likely to
be observed, it removes all null values.  
Depending on the method and sample (Section \ref{sec:grad_method}), factoring
these probabilities in the linear fit typically increases the raw age gradient
(Table \ref{tab:ts}).

\subsubsection{Population synthesis modelling}\label{sec:psyn}

Our second method to explore target selection effects also relies on population
synthesis. However, rather than generating a probability data-cube, we produce
a synthetic population with a certain star formation history, metallicity
distribution function, IMF, and stellar density profile. This gives us the
flexibility of varying each of the input parameters at the time, to explore
their impact on a population. 
\begin{figure*}
\begin{center}
\includegraphics[scale=0.57]{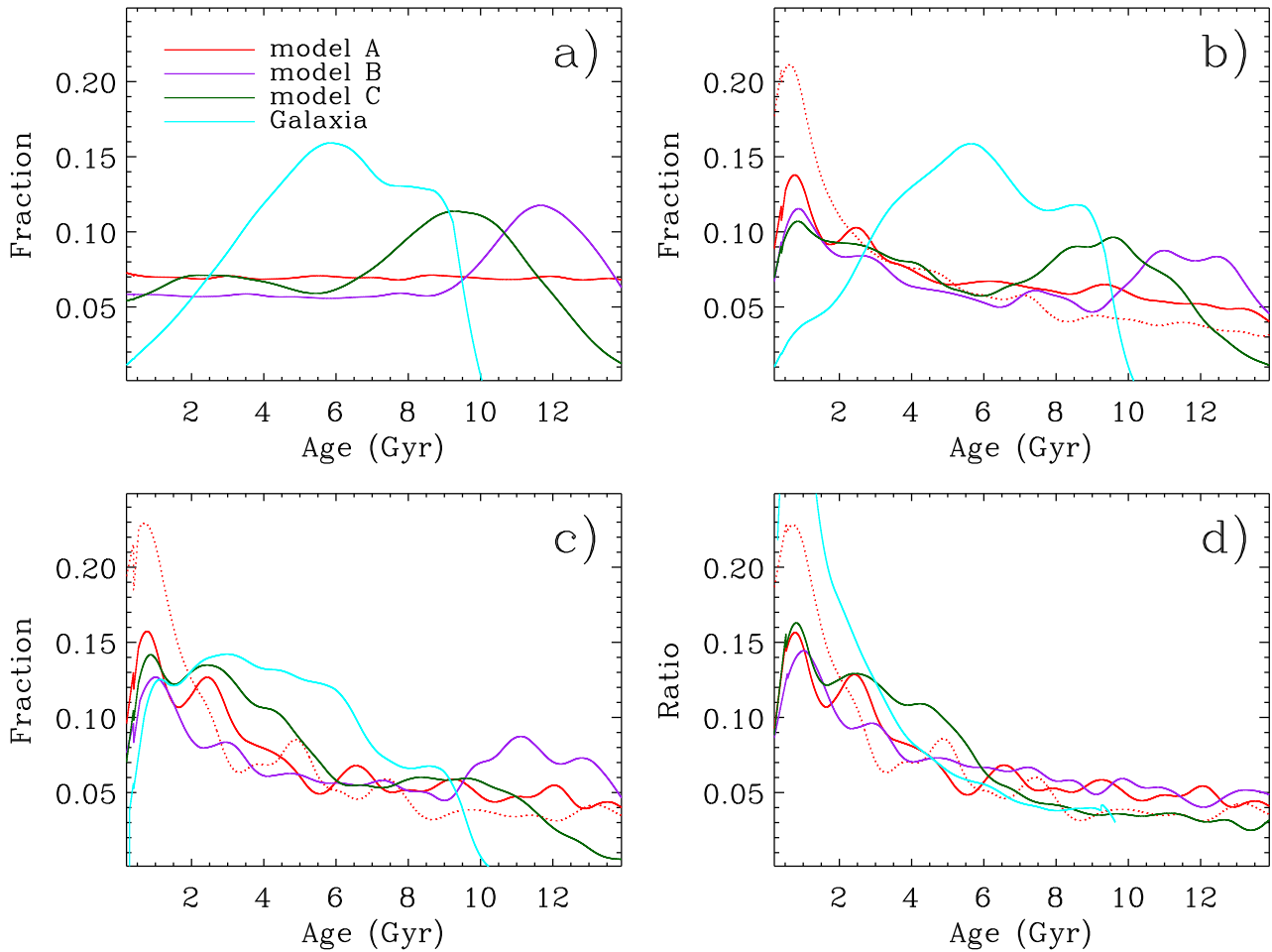}
\caption{{\it Panel} a): age distribution input in different models, as
  traced by low-mass, unevolved stars. {\it Panel} b):
  the same distributions when restricting to stars with
  $2.0 < \logg < 3.5$. Dotted red line is when changing the IMF in model A to
  have $\alpha=-1.35$. {\it Panel} c): age distributions when applying the
  SAGA ``Giants'' target selection: $2.0 < \logg < 3.5$, $10 \le y \le 13.5$,
  $0.6 \le b-y \le 0.8$ and $\feh>-1$. {\it Panel} d): ratio between the
  outputs in panel c) and the inputs in panel a).  All curves are
  normalized to equal area. Small wiggles are due to realization noise.}
\label{f:selef}
\end{center}
\end{figure*}

We assume a vertical stellar density profile described by two exponential
functions with scale-heights of $0.3$ and $1.2$~kpc, to mimic the thin and the
thick disc, respectively. For our tests, we define three models; in our first
one (A) we adopt a constant star
formation history over cosmic time, meaning that each age has a probability of
occurring $1/\tau_{max}$, where $\tau_{max}$ is the maximum age covered by the
isochrones. We also assume a flat metallicity distribution function over the
entire range of the BaSTI isochrones and a Salpeter IMF. Shallower and
a steeper slopes for the IMF are also explored ($\alpha \pm 1$).

Our second model (B) is very similar to the previous one, the only difference
being a burst of star formation centred at 12 Gyr (50 per cent of the stars),
followed by a flat age distribution until the present day. Note that in both
model A and B ages are assigned independently of their thin or thick disc
membership, and no vertical age gradient is present. 

In our last model (C) we describe the ages of thin disc stars with a
standard gamma distribution (with $\gamma=2$) having a dispersion of $2.5$~Gyr,
centred at zero on the Galactic plane, and with a vertical gradient of
$4.5$~Gyr kpc$^{-1}$. For thick disc stars we adopt a Gaussian distribution
centred at 10~Gyr with a dispersion of $2$~Gyr. The metallicity distribution
function of thin disc stars is modelled by a Gaussian centred at solar
metallicity on the plane, with a dispersion of $0.2$~dex and a vertical
gradient of $-0.2$ dex kpc$^{-1}$. For the thick disc we assume a Gaussian
metallicity distribution centred at $\feh=-0.5$ with dispersion of $0.25$~dex.
While model C provides a phenomenological description of some of the features
we observe in the Milky Way disc, it is far from being a complete representation
of it, which is not our goal anyway. A more complete Milky Way model is
explored in the next Section using Galaxia \citep{sharma11}.

Here, we simply want to explore selection effects, in particular on stellar
ages. Figure \ref{f:selef} shows how the age distribution input in different
models (traced by unevolved low mass stars, panel {\it a}) is altered when
selecting evolved stars (defined as having $2.0 < \logg < 3.5$, panel {\it b})
or applying the SAGA ``Giants'' target selection discussed in Section
\ref{sec:ss} (panel {\it c}).
It is clear that even in the simplest case (model A), the age distribution of
``Giants'' is strongly biased towards young stars (panel {\it d}), in
agreement to what we already deduced from Figure \ref{f:nifty}. This is
driven by the combined effect of evolutionary timescales and the slope of the
IMF (compare continuous and dotted line for model A).

We first compute the gradient input in each model using its unevolved
stars\footnote{Any Salpeter-like IMF breaks at sub-solar mass \citep[e.g.][and
    references therein]{bcm10}. However, this will only change the
  density of low mass stars, but not the underlying age structure they trace.},
defined here as all stars with masses below $0.7 M_{\odot}$. Because of
the large dispersion of ages at each height (also present in Galaxia,
see next Section), we find that fitting heights as function of ages --i.e.~to
derive a slope in kpc Gyr$^{-1}$-- provides a better description of the data.
From the population synthesis volume, we extract a pencil beam with Galactic
latitudes 
$8^{\circ}<b<20^{\circ}$, apply the target selection of ``Giants'' and ``RGB''
stars and compute the kpc Gyr$^{-1}$ slopes of these sub-samples. The change in
slope between the unevolved-stars and the target-selected ones defines the
correction that must be applied to the raw data. Thus, we use the correction
in slope determined above to modify the observed SAGA values, by adjusting the
height of each star depending on its age. 
This adjustment increasingly affects older stars, which are lifted in
altitude Z after correcting for target selection. In reality, the position of
each of our targets is well determined (within its observational
uncertainties): the change we introduce here is simply meant to counteract the
bias introduced into the distribution by target selection. This 
is to say that if our sample were not affected by target selection, we 
would preferentially observe additional stars at higher Z.
Once we adjust the height of each of our objects as described, we then 
perform a least-squares fit on the shifted points in terms of Gyr kpc$^{-1}$. 
We apply a similar technique to our boxcar-smoothing analysis except here, 
rather than shifting every star, we shift each median point and re-fit them 
with a least-squares in Gyr kpc$^{-1}$. Thus, although we apply the same target 
selection correction, its effect will be different. Because there is a much
wider range of values star-by-star than in the median points, the gradient
from the least-squares analysis changes more than for the boxcar-smoothing.
Also in this case, correcting for target selection typically increase the raw
SAGA gradients by a few Gyr kpc$^{-1}$ (Table \ref{tab:ts}).

\subsubsection{Galaxy modelling}\label{sec:galaxia}

By applying our target selection criteria to a model of the Galaxy, we can
determine how well the resulting sample reflects the disc behaviour assumed by
the model. For this purpose we simulate the SAGA stripe using Galaxia
\citep[][]{sharma11}.

Galaxia is based on the Besancon analytical model of the Milky Way
\citep{robin03}; the disc is composed of six different populations with a
range of ages from 0 to 10\,Gyr. The thick disc and halo are modelled as
single-burst, metal-poor populations of 11 and 14\,Gyr, respectively. For our
analysis, we limit ourselves to the six thin-disc populations in Galaxia; this
age range is representative of the bulk of the SAGA sample with a more
continuous distribution in age and chemistry than if we used also the
single-burst populations. Although the origin of the thick disc is still
unclear, it is unlikely to consist of stars having a single age and it might
also span a large metallicity range (see discussion in the Introduction).
Galaxia itself is a sophisticated --yet simplified-- representation of the
Galaxy, which assumes a certain age and metallicity distribution for each
Galactic component. Among other things the metallicity scale, the stellar
radii, gravities, synthetic colours, model $\teff$ along the red giant branch
and mass-loss prescription will also depend on the isochrones implemented in
the model, which are from Padova in the case of Galaxia
\citep{bertelli94,marigo08}. We do not attempt to vary any of the Galaxia
ingredients, and we have already explored the effect of changing some of those
assumptions using the population synthesis approach described in the previous
Section.

Here, we want to further assess how a known input population from a realistic
Galactic model will appear once filtered through our target
selection algorithm. We adopt the same technique described in Section
\ref{sec:psyn}. We calculate the input gradient using unevolved stars, implement
the ``Giants'' and ``RGB'' target selection on the Galaxia simulated stars to
derive corrections in kpc Gyr$^{-1}$, and apply those to the data before
re-fitting the gradient in Gyr kpc$^{-1}$. The age distribution input in
Galaxia is rather different from that traced by our simplistic population
synthesis models, and it does not extend beyond 10~Gyr because of the thick
disc exclusion (Figure \ref{f:selef}). 

The Galaxia model shows a wide range of ages at each height above the Galactic 
plane; however, the proportion of young stars diminishes as the height 
increases, resulting in typically older ages far away from the Galactic plane.
The SAGA cuts in colour and magnitude remove many of the older stars at large
heights: this boosts the fraction of young stars and skews the sample to lower
heights in accordance to what we already derived in Section \ref{sec:nifty} and
\ref{sec:psyn}. Target selection corrections are similar to what we
derived previously, and of the order of few Gyr kpc$^{-1}$.

\subsubsection{Correlation with distances}
\begin{figure*}
\begin{center}
\includegraphics[scale=0.1]{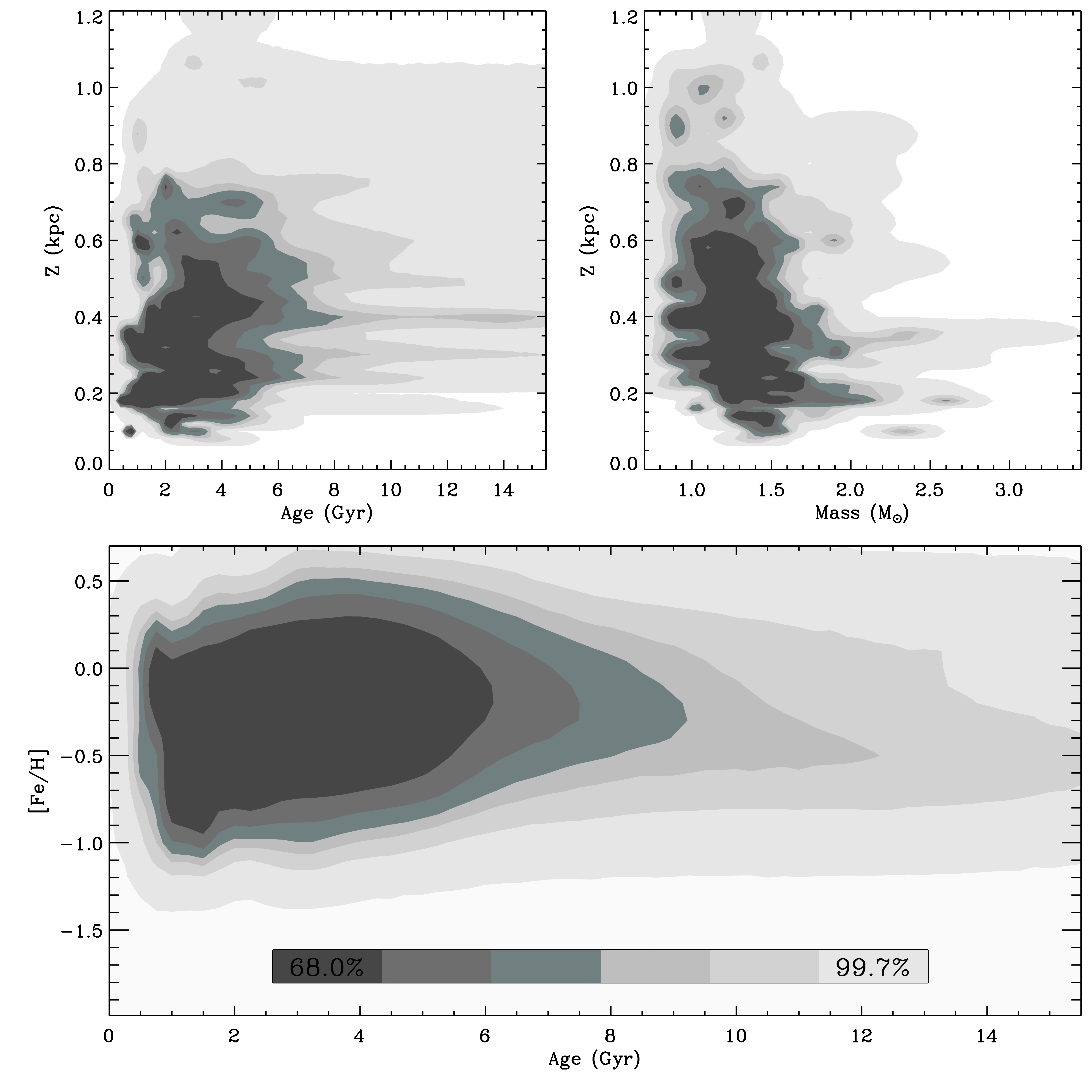}
  \caption{Age and mass gradients ({\it top panels}) and age-metallicity
  relation ({\it lower panel}) after correcting for target selection the
  ``Giants'' sample. Contour levels have been obtained by convolving each
  star with its age, distance and metallicity uncertainties and assigning a
  density proportional to the logarithm of the inverse probability of being
  observed. Probabilities have been computed as described in Section
  \ref{sec:nifty}.}
\label{f:wh2}
\end{center}
\end{figure*}

In a pencil-beam sample such as SAGA, the average altitude $\textrm{Z} = D
\sin(b)$ will, by the 
geometry, rise almost linearly with distance $D$;  hence the two quantities 
are strongly correlated. Thus, any correlation for example of age with
distance, will bias the gradient derived as function of Z. This effect can be
accounted for by introducing the dependence on distance in the least-squares
fit when deriving the gradients \citep[e.g.][]{sac14}. This technique provides
a model-independent check (modulo the
degree at which giants trace the ages of an underlying stellar population).
Assuming that a (multi) linear dependence provides a reasonable description of
the underlying structure of the data (which over the range of distances
studied here is appropriate for ages), one can expand the fit into
\begin{equation}\label{eq:ovb}
\tau_i = \frac{d \tau}{d\textrm{Z}} \textrm{Z}_i + \frac{d \tau}{dD} D_i + \epsilon
\end{equation}
where $i$ is the index running over the stellar sample, $d\tau/d\rm{Z}$ and 
$d\tau/dD$ are the free fit parameters measuring the correlation between 
age $\tau$, altitude Z and distance $D$, and $\epsilon$ is the 
intercept of the fit. When we apply this technique to SAGA, the significance 
of the derived slopes is usually above three sigma for the ``Giants'' sample, 
whereas it is below 1 sigma for ``RGB'' stars due to the smaller sample size
and range of distances. Thus, we apply this method only to ``Giants''. 

Accounting for the distance dependence returns a least-squares gradient of
$6.3 \pm 1.6$ Gyr kpc$^{-1}$.
The increase with respect to the value of $4.1 \pm 0.9$ Gyr kpc$^{-1}$
obtained with a simple linear fit (Section \ref{sec:grad_method}) tells us
that the survey geometry is indeed biased against old stars, and thus any fit
of the raw data underestimate the true age gradient. 

\subsection{The vertical age gradient}\label{sec:ag} 

In Section \ref{sec:grad_method} we have used two different methods and samples
to measure the raw vertical age gradient with SAGA. We have then assessed
target selection effects using different approaches. Although they return a
range of values for the correction, they all consistently show that
any raw measurement of the vertical age gradient using red giants
underestimates the real underlying value.

We summarize the raw gradients obtained using different samples and methods in
Table \ref{tab:ts}, along with the target selection corrections discussed in
Section \ref{sec:nifty} to \ref{sec:galaxia}. For each method and sample listed
in the table, we compute the median target selection correction and standard
deviation as a measure of its uncertainty. This is added in quadrature to the
undertainty derived for each fit, after which the weighted average of all
gradients is computed, obtainining a value of $4.3 \pm 1.6$ Gyr kpc$^{-1}$.
If we instead replace the ``RGB'' slopes with those obtained without forcing
the intercept, then we obtain a weighted average of $3.9 \pm 2.5$ Gyr
kpc$^{-1}$. Hence, the gradient does not change dramatically, but its
uncertainty is increased.

\begin{table*}
\centering
\caption{Target selection effects. Corrections are intended to be summed to
  the raw vertical gradients. All values are in Gyr kpc$^{-1}$.}\label{tab:ts}
\begin{tabular}{cccccccc}
\hline
               &                 & Corrections from           & \multicolumn{3}{c}{Corrections from}                 & Corrections from   &            \\
               & Raw gradients   & target selection           & \multicolumn{3}{c}{population synthesis}             & Galactic           &            \\
               &                 &    modelling               & \multicolumn{3}{c}{modelling}                        & modelling          &            \\
               &                 &                         & A & B & C  &    &   \\
\hline
 boxcar        &  $+3.9\pm1.1$   &          $+0.6$            & $+0.2$ & $+0.3$ & $+0.1$                             & $+0.0$             & ``Giants'' \\ 
 smoothing     &  $+3.6\pm1.7$   &          $+1.2$            & $+1.4$ & $+1.2$ & $+3.5$                             & $+3.4$             & ``RGB''    \\
               &           &                            &        &        &                                    &                    &            \\ 
 least-squares &  $+4.1\pm0.9$   &          $+3.7$            & $+2.4$ & $+1.2$ & $+2.7$                             & $+3.8$             & ``Giants'' \\
  fit          &  $+1.4\pm1.2$   &          $-1.0$            & $+1.8$ & $+1.3$ & $+5.3$                             & $+4.8$             & ``RGB''    \\ 
 \hline
\end{tabular}
\begin{minipage}{1\textwidth}
  The raw gradients quoted for the ``RGB'' sample are obtained forcing the
  intercept of the fit. If kept unconstrained, the raw ``RGB'' values would
  change to $-0.1\pm3.3$ and $0\pm1.7$ Gyr kpc$^{-1}$.
\end{minipage}
\end{table*}

While all the above values clearly indicate that the age of the Galactic disc
increases when moving away from the plane, the consistency among different
samples, methods and target selection corrections vary. It should
also be kept in mind that mass-loss changes our age estimates. If we were to 
adopt the ages derived for SAGA assuming an efficient mass-loss ($\eta=0.4$),
the raw gradients for the ``Giants'' sample would decrease by $1.3$ Gyr
kpc$^{-1}$. Since the effect of mass-loss for the SAGA ``RGB'' stars is
negligible, their gradient decreases by only $0.2$\,Gyr kpc$^{-1}$.

Based on the above discussion, we conclude that in the region of the Galactic
disc probed by our sample, the vertical age gradient is on the order of
$4.0 \pm 2.$ Gyr kpc$^{-1}$, which also encompasses the uncertainty stemming
from mass-loss. In particular, it should be stressed that at any given height
there is a wide range of ages. Figure \ref{f:wh2} shows such
overdensities in the vertical age, mass and age-metallicity relation when
including observational uncertainties and correcting for target selection.

To our knowledge, the present study is the first of this kind, quantifying the 
in situ vertical age gradient of the Milky Way disc. While the origin
  of this age gradient is beyond the scope of this paper, its existence has
long been known by indirect evidence such as e.g.~the
age-velocity dispersion relation \citep[e.g.][]{vH60,mayor74,w77,holmberg07},
the chemistry in red giants \citep{mg15} and the change in fraction of active M
dwarfs of similar spectral type at increasing Galactic latitudes
\citep[e.g.,][and references therein]{west11}. However none of these studies
is able to provide a direct measurement as we do here.

\section{The age-metallicity relation of disc red giants and their age
  distribution}\label{sec:agemet}

An important constraint for Galactic models is provided by the time 
evolution of the metal enrichment, the so-called age-metallicity relation. The 
strength or even the existence of this relation among disc stars has been 
largely debated in the literature because of the intrinsic difficulty of 
deriving reliable ages for field stars, as well as issues with sample
selection biases 
\citep[e.g.,][]{mct76,twarog80,edvardsson93,ngb98,rp00,fhh01}. We can now take 
a fresh look at this issue, with the first age-metallicity relation from 
seismology shown in Figure \ref{f:wholehog} for the entire dataset, as 
well as when restricting only to ``Giants''. The SAGA target selection
intrinsically favours metal-rich, young stars thus flattening the overall
age-metallicity. If we adopt the age ($\simeq 4 \pm 2$\,Gyr\,kpc$^{-1}$) and
metallicity gradients \citep[$\simeq -0.2\pm0.1$\,dex\,kpc$^{-1}$,][]{sch14}
measured over the SAGA stripe we obtain a shallow slope of
$-0.05 \pm 0.06$\,dex\,Gyr$^{-1}$. This is consistent with what is obtained
instead if we were fitting the age-metallicity in Figure \ref{f:wholehog}, and
correcting for target selection afterwards.  
Seismology thus confirm the rather mild slope and large spread at all ages 
in the age-metallicity relation of disc stars, as already derived from 
turn-off and subgiant stars in the solar neighbourhood 
\citep[e.g.,][]{nordstrom04,haywood08,c11,brs14} and also in agreement with 
the study of Galactic open clusters \citep[e.g.,][see also \citealt{lvm13} for  
the age-metallicity relation of disc globular clusters]{friel95,cnp98}. It
should also be noted that a typical age uncertainty of order 20 per cent
implies a much larger absolute number at older ages than at younger ones (i.e.
$10\pm2$ Gyr versus $1\pm0.2$ Gyr). Thus, despite old and metal-rich stars do
exist, when convolving their uncertainties in the age-metallicity relation
their contribution is much reduced (compare Figure \ref{f:wholehog} with
Figure \ref{f:wh2}). Also, our sample selection limits us to $\feh>-1$,
preventing us from tracing the early enrichment expected in the
age-metallicity relation (compare e.g.~the steep rise in metallicity at about
13~Gyr in figure 16 of \citealt{c11}).

\begin{figure}
\begin{center}
  \includegraphics[scale=0.27]{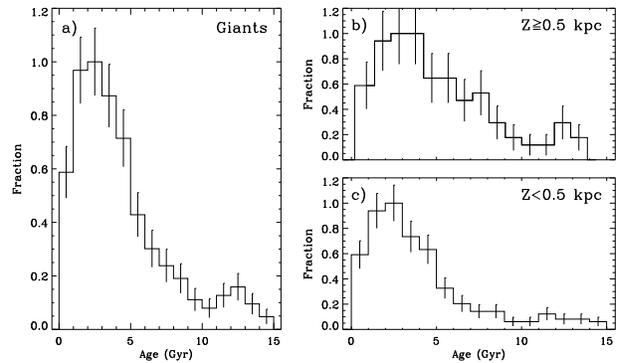}
\caption{{\it Panel} a): age distribution (with Poisson error bars) for the 
``Giants'' sample. {\it Panel} b) and c): same as above, but for 
  $\rm{Z} \ge 0.5$~kpc and $\rm{Z} < 0.5$~kpc, respectively.}
\label{f:agedist}
\end{center}
\end{figure}

Figure \ref{f:agedist} shows the age distribution for the ``Giants'' sample. 
Overall this is similar to what we have already discussed in Section 
\ref{sec:ages}, 
apart from the fact that we are now applying completness cuts. A significant
overdensity seems to appear at the oldest ages, above $\simeq 10$\,Gyr,
which persist also when adopting ages computed with mass-loss. We know 
that our target selection is biased against old stars (Section
\ref{sec:grad_bias}), and it would thus be intriguing to interpret this
overdensity as the signature of a population formed/accreated early in the
history of the Galaxy. As we have discussed in Section \ref{sec:psyn}, a
constant star formation rate produces an age distribution of red giants which
peaks at young values, and with a long tail. A strong burst in star formation
at a given age manifests instead as a localised peak at that epoch (see
Figure \ref{f:selef}).

We only select stars with $\feh>-1$, implying that this overdensity is
associated with disc stars, rather than the halo, and it could be the
signature e.g., associated to the formation of the thick disc or enhanced star
formation in the early Galaxy \citep[c.f.][]{hay13,robin14,sh14}. Because of
the vertical age gradient and the survey geometry we must first 
verify whether this overdensity could simply stem from stars 
at the highest Z. Correcting the histogram for the vertical age gradient is not 
straightforward since we have a mixture of young and old stars at all heights, 
and this would unphysically shift part of the age histogram at negative values.
We therefore split the age distribution below and above $\textrm{Z}=0.5$~kpc
in Figure \ref{f:agedist}b) and c). A moderate overdensity at the oldest ages
is still present in both panels. However, when we fold age uncertainties in
the histogram, the overdensity at old ages disappears, consistently with the
lower panel in Figure \ref{f:wh2}. 

We thus conclude that the detection of a peak at old ages is not significant
and emphasize the importance of taking proper age uncertainties into
consideration when conducting this kind of analysis. Future larger datasets
with improved age precision will be able to look for the existence of
signatures of this kind. With our current SAGA sample, we can rule out the 
presence of any major overdensity at ages younger than about 10~Gyr, implying 
that the Milky Way disc had a relatively quiescent evolution since a redshift
of about 2 \citep[see also][]{greg15}.
Increasingly sophisticated cosmological simulations are now able to predict
gross morphological properties on galactic scales \citep{torrey12,vgs14}, yet
the survival of discs seem to critically depend on the abscence of violent
events \citep{scanna09}; our results support such scenario.

\begin{figure*}
\begin{center}
\includegraphics[scale=0.6]{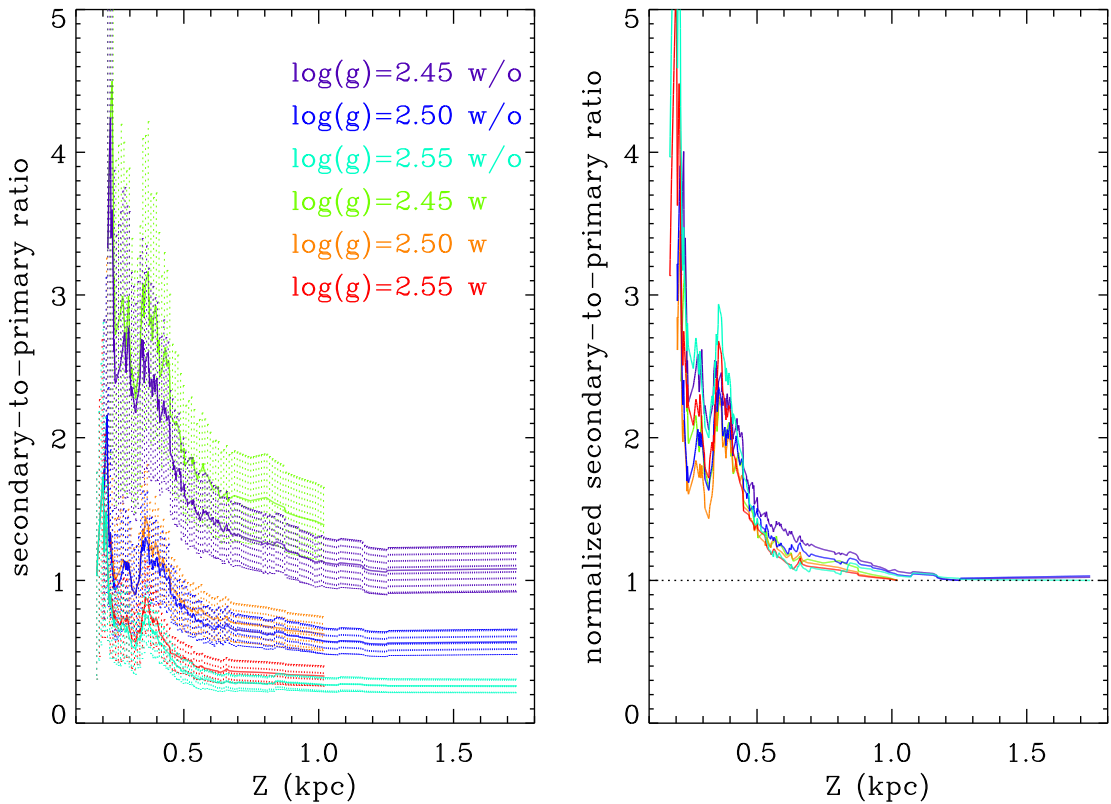}
\caption{{\it Left panel:} ratio between the (cumulative) number of secondary 
to primary clump stars and function of height from the plane Z. Different 
colours indicate the adopted surface gravity cut to discriminate between
primary 
and secondary clump, whether with (w) or without (w/o) completness cuts. Dotted 
areas indicate Poisson's errors. {\it Right panel:} same plot but normalizing 
each curve to the value at the highest altitude Z.}
\label{f:SC}
\end{center}
\end{figure*}

\section{Secondary clump stars: standard age candles for Galactic Archaeology}\label{sec:rc2}

The secondary clump is populated by stars which ignite helium in (partly) 
nondegenerate conditions, and it is a phase relatively well-defined in time
\citep[e.g.,][]{girardi99}. 
Although the precise mass and hence age, at which this happens depend on the 
models themselves, secondary clump stars define a nearly pure population of 
young ($\lesssim 2$~Gyr) stars. At the youngest ages the intrinsic luminosity
of clump stars is non constant, thus making them unsuitable distance
calibrators (Chen et al., in prep); nevertheless as we show below, secondary
clump stars can be used as standard age candles to estimate {\it i)} the
intrinsic metallicity spread at young ages, and {\it ii)} to trace the aging
of the Galactic disc.

{\it i)} We have already shown that thanks to our precise seismic $\logg$
determinations
we can discriminate between primary and secondary clump stars in a field
population (see also Paper I, figure 17). Here, we use only stars with secure
``RC'' classification and adopt a fixed $\logg=2.5$ to discriminate between
the two phases. This cut is rather arbitrary, although we have verified that
our results still hold for reasonable variations of this threshold. After
discarding members 
of the open cluster NGC\,6819 and considering only stars with good photometric 
and metallicity flags, we derive for the secondary clump stars a metallicity 
scatter $\sigma_{\feh}=0.28 \pm 0.04$~dex. 
The  reported uncertainty is twice as large as the variation stemming from a
change of $0.05$~dex in gravity cut, and from the adoption or not of 
the completeness cuts ($0.6 \le b-y \le 0.8$ and $y \le 13.5$ as derived in 
Section \ref{sec:stat}). After unfolding the typical uncertainty of our
photometric metallicities (derived by comparing with independent $\feh$
measurements, Paper I), we find that the intrinsic metallicity spread of
secondary clump stars is $0.14 \pm 0.04$~dex (this procedure holds true
  under the assumption that errors are reasonably Gaussian, and there is no
systematic bias). 
This number is essentially unchanged when correcting for the vertical 
metallicity gradient measured in SAGA \citep{sch14}. The intrinsic scatter 
derived here is similar to that obtained using stars in the same age range
from the GCS \citep{c11}, which, after accounting for the 
uncertainty in those metallicity measurements, is about $0.13 \pm 0.02$~dex. 

{\it ii)} Since primary and secondary clump stars occupy very similar position
on the H-R diagram, it is reasonable to assume that they have 
nearly equal probability of being observed by {\it Kepler}. Therefore, the 
ratio of secondary to primary clump stars is independent of the selection 
function. Most importantly, this ratio is sensitive to the mixture between a 
young population (including secondary clumps) and an old one (including 
primary only), thus meaning that it can be used to trace the relative age 
of a population. Also, although ages in the clump phase are affected by 
mass-loss (especially at the low masses typical of primary clump stars) 
their number ratio is unaffected by this uncertainty, until linked to an age 
scale. 

This is explored in Figure \ref{f:SC} (left-hand panel), which shows the ratio
of secondary to primary clump stars as function of height from the Galactic
plane Z. Again, members of the cluster NGC\,6819 have been excluded since we
are interested at studying properties of field disc stars.
Adopting our completness cuts ($V \le 13.5$ and $0.6 \le b-y \le 0.8$) is 
irrelevant here, with little effect aside from changing the limit in heights 
at which we have targets.
The ratio of secondary to primary clump stars will vary when adopting
different cuts in $\logg$ ($2.45$, $2.50$ and $2.55$~dex). However, once these
ratios are normalized to the value at the maximum height, they all remarkably
overlap. It is clear that at lower altitudes secondary clump stars outnumber 
primary ones, indicating that the fraction of young stars decreases when
moving away from the plane, in qualitative agreement with the vertical age
gradient measured in Section \ref{sec:ag}.

\section{Conclusions}\label{sec:end}

In this paper we have used the powerful combination of asteroseismic and
classic stellar parameters of the SAGA ensemble to investigate the vertical
age structure of the Galactic disc as traced by red giants in the \kepler field.
This goal is facilitated by the 
pencil-beam survey geometry analyzed here, which covers latitudes from about 
$8^{\circ}$ to $20^{\circ}$ translating into bulk vertical distances up to 
$\approx 1$~kpc above the Galactic plane for red giants. Galactic longitudes 
are centred at 
$l \simeq 74^{\circ}$ implying nearly constant Galactocentric distances and 
thus minimizing radial variations. 

For the asteroseismic sample we have complemented the stellar masses,
metallicities and distances already derived in Paper I with stellar ages. For
a large fraction of our 
stars we have seismic classification available to distinguish between red 
giants burning hydrogen in a shell and clump stars that have already ignited 
helium in their core, thus greatly improving on the accuracy of age 
determinations. For clump stars, as well as for stars on the upper part 
of the red giant branch the largest source of uncertainty in age determination 
stems from mass-loss. We have therefore included this uncertainty by deriving
stellar ages under two very different assumptions for mass-loss.

The \stromgren photometry of SAGA is magnitude complete to $y \approx 16$, 
i.e.~nearly two magnitudes fainter than the giants selected to measure stellar
oscillations with the \kepler satellite. This, and the capability of
Str\"omgren photometry to disentangle dwarfs from giants, has allowed us to
build an unbiased population of giants, that we have used to benchmark against
the asteroseismic 
sample. We have been able to constrain the thus-far unknown selection 
function of seismic targets for the \kepler satellite (see also Sharma et al.,
in prep.), by identifying a colour 
and magnitude range where giants with oscillations measured by \kepler are 
representative of the underlying population in the field. This holds true for 
$V=y\le 13.5$ and $0.6 \le b-y \le 0.8$, modulo reddening, which is anyway well
constrained for our sample. This has been verified to 
correspond to $K_S \le 13 - 3.28 (J-K_S)$ and $0.586 \le J-K_S \le 0.782$ for 
the 2MASS system. 
These cuts, together with the use of stars with best quality flags in SAGA, 
as well as the exclusion of members of the open cluster NGC\,6819 and a 
handful of the most metal-poor stars ($\feh \le -1$, for which seismic scaling 
relation might be inaccurate) reduce the initial SAGA sample by almost one 
third, to 373 stars. 

Although we have been able to identify the colour and magnitude range where our 
sample is representative of giants in the field, when measuring the vertical 
age structure of the Galactic disc we must still correct the raw measurements 
for the colour and magnitude cuts reported above, i.e. for target selection. 
To control for these biases, we separately estimated the effects of the
selection function from Galaxy models, and from a more simple and
straightforward approach with direct population synthesis.

We see a clear increase of the average stellar age at increasing Galactic 
heights, thus indicating the aging of the Milky Way disc as one moves away from 
the Galactic plane. This is also traced by a stellar mass gradient, since the
mass of a red giant is a proxy for its age. We have used linear fits to
describe 
these trends; although this allows us to quantify their strength, we are aware 
that they might not capture the full complexity of the age and mass structure 
in the Galactic disc. The bulk stellar age increases with increasing altitude, 
but there is a large spread of ages at all heights. This translates into a 
decreasing stellar mass with increasing altitude; stellar masses are not 
linearly mapped into ages, and the overall trend of the stellar mass with 
Galactic heights is rather L-shaped. 

We have quantified these trends using giants independently of their seismic 
classification (373 stars), as well as ``RGB'' stars only (48 object), for 
which the impact of mass-loss on age estimates is negligible. All the above 
estimators and samples agree in showing increasing stellar ages (and
decreasing stellar masses) at increasing Galactic heights, albeit the degree
of consistency among different methods and samples varies. We have argued that
our current best estimate for the vertical age gradient is $4 \pm 2$\,Gyr
kpc$^{-1}$. Part of the scatter might stem from uncertainties related to
sample size and target selection corrections, although it should also be kept
in mind that part of it is real, an the age gradient we measure is just the
heighest overdensity of a wide distribution.
We have also used the number ratio of secondary clump stars to primary clump
stars as an independent proxy of the aging of the stellar disc, confirming the
presence of preferentially old stars at increasing Galactic heights. 

Stellar ages show a smooth distribution over the last 10 Gyr, whereas a
small overdensity appears at older values, which could be a signature
associated with the early phases of the Milky Way. Once age uncertainties are
taken into account, this does not appear to be statistically significant.
Nevertheless, the smooth distribution of ages over the last 10~Gyr is
consistent with a rather constant star formation history and suggests that the
Galactic disc has had a rather quiescent evolution since a redshift of about 2.
This is in agreement with scenarios where stellar discs in galaxies form at
relatively early times, and their survival critically depends on the absence
of major mergers. 

Finally, we derive the first seismic age-metallicity relation for the Galactic
disc. We confirm results from other methods (such as age dating of turn-off
and subgiant stars, as well as Galactic open clusters) that a metallicity spread
exists at all ages, and the overall slope of the age-metallicity relation is
small. Because of their young ages, secondary clump stars can also be used to 
assess the instrinsic metallicity spread at almost the present time, which we
estimate to be $\approx 0.14$~dex. We remark that studies of local early type
stars and gas-phase in diffuse interstellar medium reveal indeed a high degree
of homogeneity in the present day cosmic abundances \citep{sm01,np12}.
Thus, despite a spread of ages at all heights, and a spread of metallicity at
all ages, there are well defined and smooth vertical age and metallicity
gradients, indicating that the disc is generally composed of well mixed
populations that have undergone a largely quiescent evolution. This validates
scenarios in which the evolution of the disc is largely driven by internal
dynamical processes, and it provides a first constraint on the
disc spatial growth over cosmic time. 

\section*{Acknowledgments}
We thank an anonymous referee for his/her insightful comments and suggestions
which has helped to strengthen the paper and improve the presentation of the
results. We thank P.\,E.\,Nissen and A.\,Dotter for useful discussions.
We thank the nature of who knew everything upfront for giving a good laugh.
Funding for the Stellar Astrophysics Centre is provided by The Danish National
Research Foundation (grant agreement No. DNRF106). The research is supported
by the ASTERISK project (ASTERoseismic Investigations with SONG and Kepler),
funded by the European Research Council (grant agreement No. 267864).
V.S.A. acknowledges support from VILLUM FONDEN (research grant 10118). A.M.S. is
partially supported by grants ESP2014-56003-R (MINECO), EPS2013-41268-R
(MINECO) and 2014SGR-1458 (Generalitat de Catalunya). We acknowledge the
generous hospitality of the Kavli Institute for Theoretical Physics where part
of this work was carried out. This research was supported in part by the
National Science Foundation under Grant No. NSF PHY11-25915. 

\bibliographystyle{aj}
\bibliography{refs}

\end{document}